\documentclass[prd,a4paper,longbibliography,titlepage,showkeys]{revtex4}
\usepackage{epsfig}
\usepackage{amsmath}
\usepackage{amssymb}
\usepackage{amsbsy}
\usepackage{enumerate}
\usepackage{url}
\usepackage{xspace}
\usepackage{hyperref}

\newcommand\sss{\mathchoice%
{\displaystyle}%
{\scriptstyle}%
{\scriptscriptstyle}%
{\scriptscriptstyle}%
}

\def\beq{\begin{equation}}
\def\beqn{\begin{eqnarray}}
\def\eeq{\end{equation}}
\def\eeqn{\end{eqnarray}}

\newcommand\pt{p_{\sss\rm T}}
\newcommand\pT{p_{\sss\rm T}}
\newcommand\HT{H_{\sss\rm T}}
\newcommand\kt{k_{\sss\rm T}}

\newcommand\HERWIG{{\tt HERWIG}}

\newcommand\PYTHIA{{\tt PYTHIA}\xspace}

\newcommand\sigdi{\tilde{\sigma}_2^{\sss\rm incl}}
\newcommand\sigde{\tilde{\sigma}_2^{\sss\rm excl}}
\newcommand\sigti{\tilde{\sigma}_3^{\sss\rm incl}}
\newcommand\sigte{\tilde{\sigma}_3^{\sss\rm excl}}

\def\rg{\right\}} 
\def\lg{\left\{} 
\def\({\left(} 
\def\){\right)} 
\newcommand\as{\alpha_{\sss\rm S}}

\newcommand\POWHEG{{\tt POWHEG}\xspace}
\newcommand\POWHEGBOX{{\tt POWHEG BOX}\xspace}
\newcommand{\PP}{{\tt POWHEG+PYTHIA}\xspace}
\newcommand{\PH}{{\tt POWHEG+HERWIG}\xspace}

\newcommand{\HEJ}{{\tt HEJ}\xspace}

\newlength{\wfig}
\newlength{\hfig}
\newcount\minutes 
\newcount\scratch 
 
\def\timestamp{%
\scratch=\time 
\divide\scratch by 60 
\edef\hours{\the\scratch} 
\multiply\scratch by 60 
\minutes=\time 
\advance\minutes by -\scratch 
---$\,$\hours:\null 
\ifnum\minutes< 10 0\fi 
\the\minutes}

\begin{document}

\preprint{Edinburgh 2012/01\\CP3-Origins-2012-002\\DIAS-2012-3, LPN12-032\\IPPP/12/05,
DCPT/12/10}

\title{Probing higher-order corrections in dijet production\\ at the LHC}

\author{Simone Alioli}
\affiliation{Ernest Orlando Lawrence Berkeley National Laboratory, University of California, Berkeley, CA 94720, USA}
\email[Email: ]{salioli@lbl.gov}
\author{Jeppe R.~Andersen}
\affiliation{$CP^3$-Origins, University of Southern Denmark, Campusvej 55, \\DK-5230 Odense
  M, Denmark}
\email[Email: ]{jeppe.andersen@cern.ch}
\author{Carlo Oleari}
\affiliation{Universit\`a di Milano-Bicocca and INFN, Sezione di Milano-Bicocca\\
  Piazza della Scienza 3, 20126 Milan, Italy}
\email[Email: ]{carlo.oleari@mib.infn.it}
\author{Emanuele Re}
\affiliation{IPPP, Department of Physics,
  University of Durham, Durham, DH1 3LE, UK}
\email[Email: ]{emanuele.re@durham.ac.uk}
\author{Jennifer M.~Smillie}
\affiliation{School of Physics and Astronomy, University of Edinburgh,  Mayfield Road,  Edinburgh EH9 3JZ, UK}
\email[Email: ]{j.m.smillie@ed.ac.uk}

\begin{abstract}
Both the ATLAS and CMS Collaborations have sought for effects beyond pure
next-to-leading order in dijet observables, with the goal to distinguish
between the perturbative descriptions provided by a next-to-leading order
plus collinear-resummation calculation and by the resummation of wide-angle,
hard emissions.  In this paper we identify regions of phase space in dijet
production where some observables receive large corrections beyond
next-to-leading order and study their theoretical description with two tools
that perform these two different resummations: the \POWHEGBOX and \HEJ
. Furthermore, we suggest analyses where the predictions from \POWHEG and
\HEJ can be clearly distinguished experimentally.
\end{abstract}


\maketitle

\section{Introduction}
\label{sec:introduction}
Dijet production is one of the cornerstone processes at the LHC.  The cross
section for jet production is very large, making it an important testing
ground for our understanding of QCD at high-energy scales.  In addition, jet
production is an important background for many searches for new physics. It
is therefore essential to probe and test our theoretical predictions.  

A central question is whether a framework based on a (possibly
next-to-leading-order-matched) parton shower (which resums the radiation
resulting from a large ratio in transverse scales) is sufficient for the
description of additional jets, or whether BFKL-type~\cite{Fadin:1975cb,
  Kuraev:1976ge, Kuraev:1977fs, Balitsky:1978ic} effects from hard,
wide-angle emissions, have already become important at the center of mass
energy of the LHC (7~TeV in the present study). There have been a number of
very interesting experimental studies in dijet production by both the
ATLAS~\cite{Aad:2011jz, Aad:2011tqa, Aad:2011fc} and
CMS~\cite{Chatrchyan:2011wn, Khachatryan:2011zj,Collaboration:2012gw}
Collaborations so far. From these studies it is already clear that
higher-order QCD contributions, beyond a fixed-order, low-multiplicity
calculation, can be important, because the large available phase space for
jet emission at the LHC compensates for the suppression of extra powers in
the strong coupling constant.  With the current study, we suggest analyses
which better distinguish between the two mechanisms for creating additional
jet activity: a hierarchy of transverse scales (as in the \POWHEG approach)
and the opening of phase space as the rapidity span between two jets is
increased (as implemented in \HEJ).

In this paper, we compare three theoretical approaches to dijet production: a
fixed next-to-leading order~(NLO) calculation, \POWHEG~\cite{Nason:2004rx,
  Frixione:2007vw, Alioli:2010xa} and \HEJ~\cite{Andersen:2009nu,
  Andersen:2009he, Andersen:2011hs} results.  The \POWHEG method successfully
merges a fixed next-to-leading order calculation with a parton shower
program that resums leading-logarithmic contributions from collinear
emissions. Here, the \POWHEG results obtained with the
\POWHEGBOX~\cite{Alioli:2010xd} are interfaced with the
transverse-momentum-ordered shower provided by
\PYTHIA~6.4.25~\cite{Sjostrand:2006za}.  In contrast, the starting point for
\HEJ (High Energy Jets) is an all-order approximation to the hard-scattering
matrix element in the regime of wide-angle QCD emissions.  \HEJ is accurate
at leading-logarithmic precision in the invariant mass of any two jets.  This
is then supplemented with the missing contributions (through a merging and
reweighting procedure) necessary to also ensure tree-level accuracy for final
states with up to four jets.  The tree-level matrix elements are taken from
stand-alone Madgraph~\cite{Alwall:2007st}.

Dijet production is of course important not only by itself but also when the jets are
accompanying other particles, such as the $W/Z$ vector bosons or the Higgs
boson.  For example, Higgs boson production plus two jets is an important
process in the Standard Model. 
It is known that the signature of the vector-boson fusion Higgs boson
production is two jets well separated in rapidity.
In addition, we expect very low hadronic activity between the two hardest
jets, due to the exchange of the colourless vector bosons in the $t$
channel~\cite{Dokshitzer:1991he}, contrary to what is expected for the
gluon-fusion production mechanism.  A key feature is then the study of the
efficiency of the central-jet veto, to suppress gluon-fusion
processes~\cite{DelDuca:2001eu, DelDuca:2001fn}, where it is well known that
the higher-order corrections~\cite{DelDuca:2004wt, DelDuca:2006hk,
  Campbell:2006xx, Campbell:2010cz} are very
significant~\cite{Binoth:2010ra}, and a detailed
understanding~\cite{DelDuca:2003ba, Andersen:2008ue, Andersen:2008gc,
  Andersen:2010zx} of the structure of the radiation pattern is needed.  Some
of the features in $Hjj$ production in gluon fusion are in fact
universal~\cite{Binoth:2010ra} to dijet processes in general, like $W/Z$ + 2
jets or pure dijet production, and therefore they may help as a testing
ground for techniques which can be applied in the Higgs boson searches and
studies.  Before the possible study of processes with a Higgs boson, it thus
becomes interesting to investigate the (hard) radiation pattern in events
with at least two jets, in particular for events with a non-negligible
rapidity separation between the two jets.  Recently, also the combined
effects on the cross section of a large dijet rapidity separation and a large
ratio between the transverse scale of the observed jets and a jet veto have
been investigated theoretically~\cite{Forshaw:2009fz, Schofield:2011zi,
  Delgado:2011tp}.

The layout of this paper is the following: in section~\ref{sec:phenomenology},
we discuss in detail existing experimental analyses and theoretical
predictions. Since our goal is both to investigate effects beyond NLO and to
compare results from \POWHEG and \HEJ, stable perturbative predictions for
the observables must be obtained in all three approaches. The NLO prediction
with symmetric cuts gives physically unreliable results. Therefore, in
section~\ref{sec:reliability-nlo}, we investigate various cuts that render the
NLO results reliable by removing the dependence on large unresummed
logarithmic terms, arising from soft-emission regions~\cite{Frixione:1997ks,
  Andersen:2001kt, Banfi:2003jj, Alioli:2010xa}. In
section~\ref{sec:inclusive-variables}, we propose analyses that probe the
description of radiative effects beyond NLO, and which have the potential to
better expose the differences between \POWHEG and \HEJ. Experimental
measurements of these quantities will allow a better understanding of which
is the dominant mechanism in the generation of radiation. Finally, we
summarize our findings in section~\ref{sec:conclusions}.

\section{Dijet production at the LHC}
\label{sec:phenomenology}
In this section, we discuss existing experimental analyses where
dijet data collected by ATLAS
and CMS
have been compared with the theoretical results obtained with the \POWHEGBOX and
\HEJ.  The \POWHEG and \HEJ approaches are clearly very different in their
description of QCD radiation: the former resums collinear emissions, while the
latter soft and hard, wide-angle emissions.  Nevertheless, for several kinematic
distributions (see, for example, Refs.~\cite{Aad:2011jz,Collaboration:2012gw}) the predictions from
\POWHEG and \HEJ are very similar.  This is due, as discussed below, to the
inclusiveness of the studied kinematic distributions and the specific cuts
applied in the experimental analyses. Analysis and cuts aimed at better exposing the
differences in the approaches are suggested in
Sec.~\ref{sec:inclusive-variables}.  In the distributions that we discuss in this
section, no comparison between data and the fixed NLO result is performed.

\subsection{ATLAS results}
The ATLAS Collaboration has studied the production of additional jets from a
dijet system~\cite{Aad:2011jz}.  In this study, jets are reconstructed by using the
anti-$\kt$ jet algorithm~\cite{Cacciari:2008gp} with $R=0.6$ and required to
have a transverse momentum above 20~GeV, with absolute rapidity less than 4.4.
\begin{figure}[!bt]
  \centering
  \epsfig{file=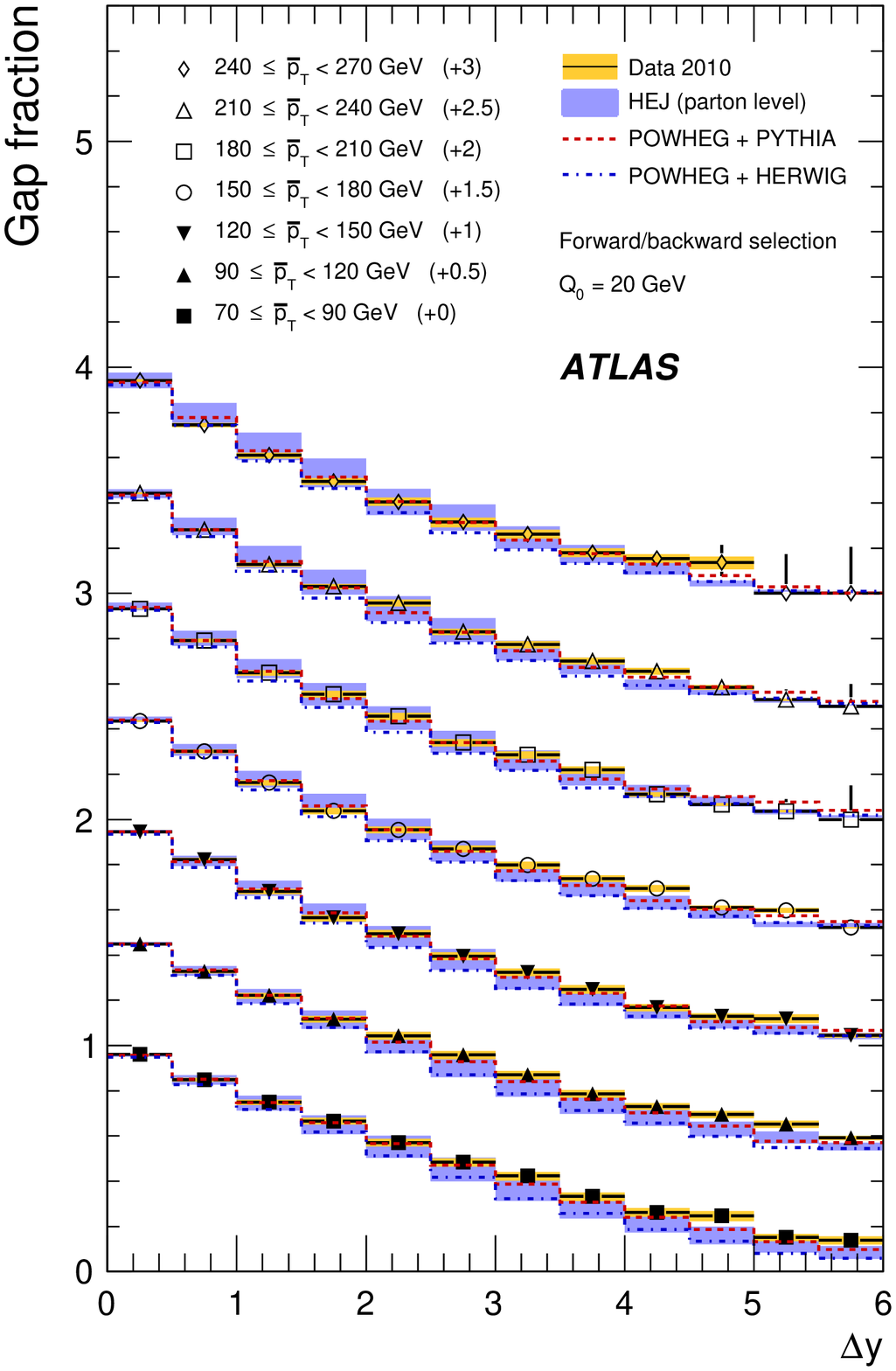,width=0.4\textwidth}
  \hspace{0.7cm}
  \epsfig{file=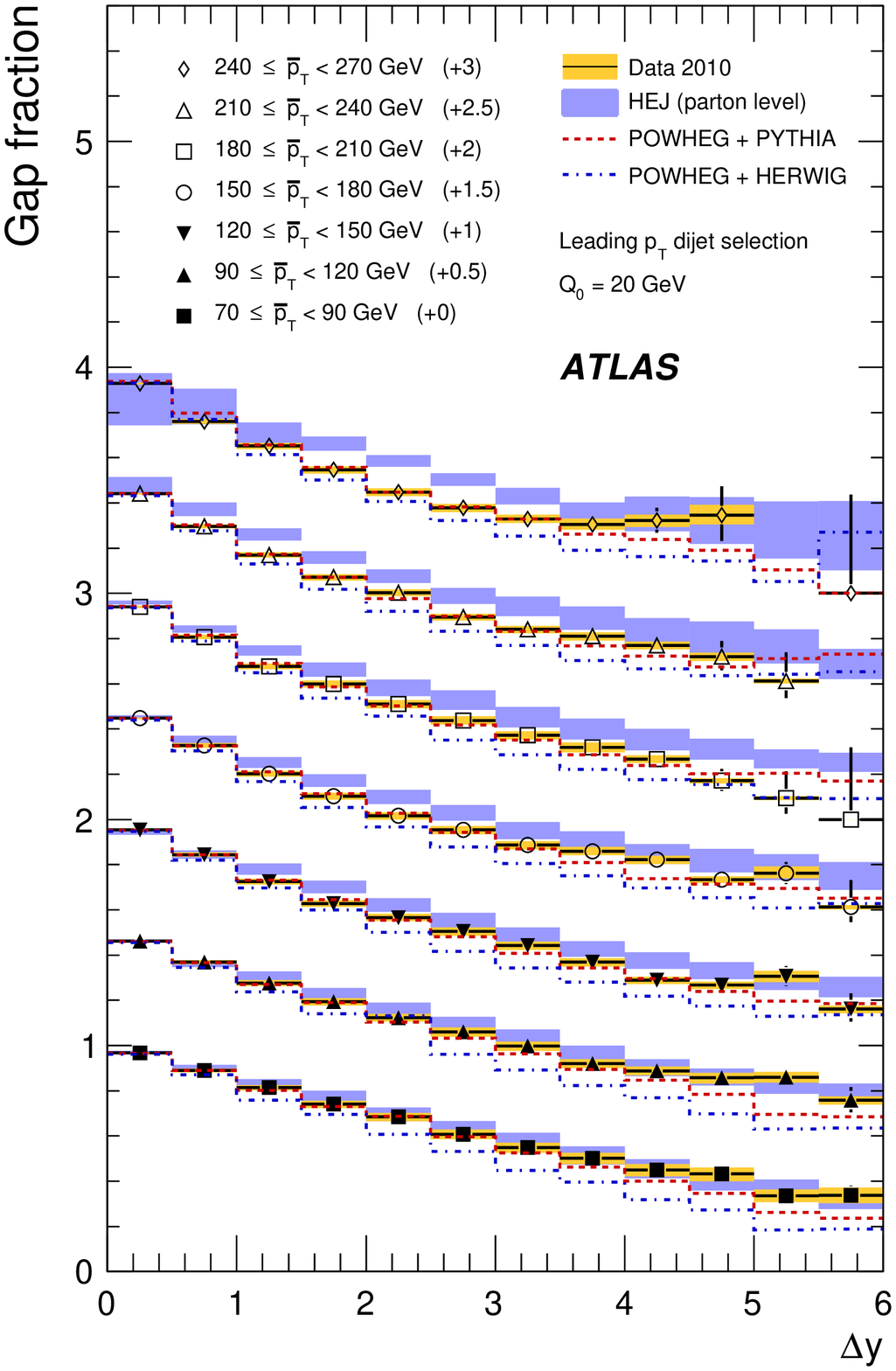,width=0.4\textwidth}
  \caption{Both plots from the ATLAS study~\cite{Aad:2011jz} on the gap
    fraction, defined as the fraction of events with no additional jets in
    the rapidity region between the tagging jets, as a function of the
    difference of the rapidity of the two jets.  In the left-hand side plot,
    the tagging jets are the most forward and most backward jet, while in the
    right-hand side plot, they are the two hardest jets in each event.  In
    both plots, $\bar{p}_{\sss T}$ is the average transverse momentum of two
    tagging jets.  All jets are required to have $\pT>20$~GeV and absolute
    rapidity $|y|<4.4$.}
  \label{fig:atlasres}
\end{figure}
We show in fig.~\ref{fig:atlasres} extracts from this ATLAS study, where the
gap fraction, defined as the fraction of events with no additional jets in
the rapidity region between the two tagging jets, is plotted as a function of
$\Delta y$, the difference of the rapidities of the two tagging jets. In the
left plot, the tagging jets are defined to be the most
forward and most backward jet (in rapidity), while, in the right plot, they
are chosen to be the two hardest jets (highest 
transverse momentum) in each event.  In both plots, $\bar{p}_{\sss T}$ is the
average transverse momentum of the two tagging jets, and results are shown for
slices of the average transverse momentum $\overline{p}_{\sss T}$ ranging
from 70 up to 500~GeV.  The experimental data are then compared with the
predictions from \HEJ [the blue band indicating the scale variation obtained
by varying the (equal) renormalization and factorization scale by a factor of
2] and the \POWHEG predictions showered by \PYTHIA and \HERWIG.

We would like to add a few comments to these findings:
\begin{enumerate}
\item 
In the left plot, the theoretical predictions obtained in the two very
different approaches of \HEJ and \POWHEG are very similar and agree with
data over a wide range of rapidity intervals and average transverse-momentum
slices.  For this kinematic quantity, where the tagging jets are selected to
be the forward/backward ones, a large hierarchy in the transverse momenta of
these jets develops as the average transverse momentum of the
forward/backward jets is increased, so that at least one of the
forward/backward jets must be very hard.  Indeed, it is observed that for
this selection, the average difference in the transverse momentum of the
forward/backward jet increases systematically with increasing $\Delta
y$. Since the veto scale ($Q_0$ in the label of the figure) for counting
additional jets is much smaller than the average transverse momentum of the
tagging jets, the jet production is driven by relatively soft emission from
the dijet system.  Both \HEJ and \POWHEG, with different approximations, do
include these multiple emissions and lead to similar results and a good
agreement with the data.

\item 
In the right plot of fig.~\ref{fig:atlasres}, where the tagging jets are chosen to
be the two hardest jets in the event, as the average $\pt$ of the two hardest
jets increases to 5 times or more of the veto scale $Q_0$, the \HEJ
prediction starts deviating from data, underestimating the amount of
radiation (i.e.~the prediction for the gap fraction is larger than the
data). This behavior is expected, since the component of events added with
na\"ive tree-level matching increases with increasing $\bar{p}_{\sss
  T}$. This component receives no systematic treatment of soft resummation
within \HEJ, a situation which would be improved by a complete matching with
a parton shower. Progress in this direction has recently been made in
ref.~\cite{Andersen:2011zd}.

The \POWHEG description includes the effects of collinear emissions through
the shower formulations, and the theoretical predictions perform well for
both kinematic distributions in fig.~\ref{fig:atlasres} (particularly when
using the \PYTHIA shower).  However, as can be seen in the right plot, for
larger rapidity spans and modest $\bar{p}_{\sss T}$, the \POWHEG description
undershoots the data.  Indeed, \POWHEG contains no systematic resummation of
all the leading-logarithmic terms for large $\Delta y$. Overall, the study
reported by ATLAS shows best agreement with the predictions of \PP, but all
the studies involve a hierarchy of transverse scales and, therefore, by
construction, will favor the description with the systematic collinear
resummation of a parton shower.

Note that the results for \PH are consistently below the data (i.e.~the
events contain too many jets). The differences between the results from \PP
and \PH should be considered as a theoretical uncertainty connected to the
different shower algorithm.
\end{enumerate}

As a final comment, to cleanly separate the two drivers of additional jet
activity (a large ratio of transverse scales and a large rapidity separation), it
is obviously necessary to use a selection criterion which does not
automatically generate a hierarchy in the transverse scales as the rapidity
span increases.

\subsection{CMS results}
\begin{figure}[!btp]
  \centering
  \epsfig{file=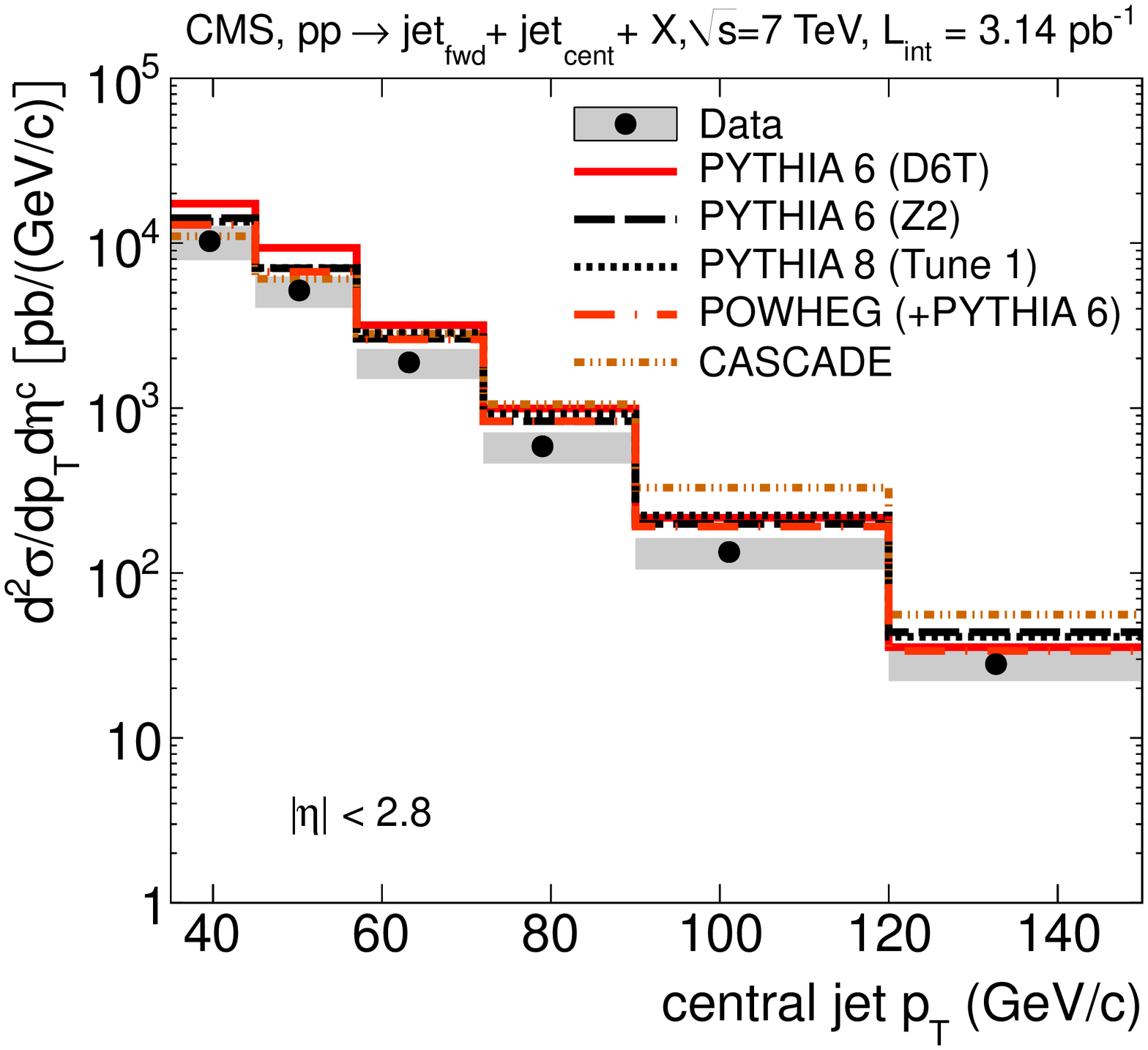,width=0.45\textwidth}\hfill
  \epsfig{file=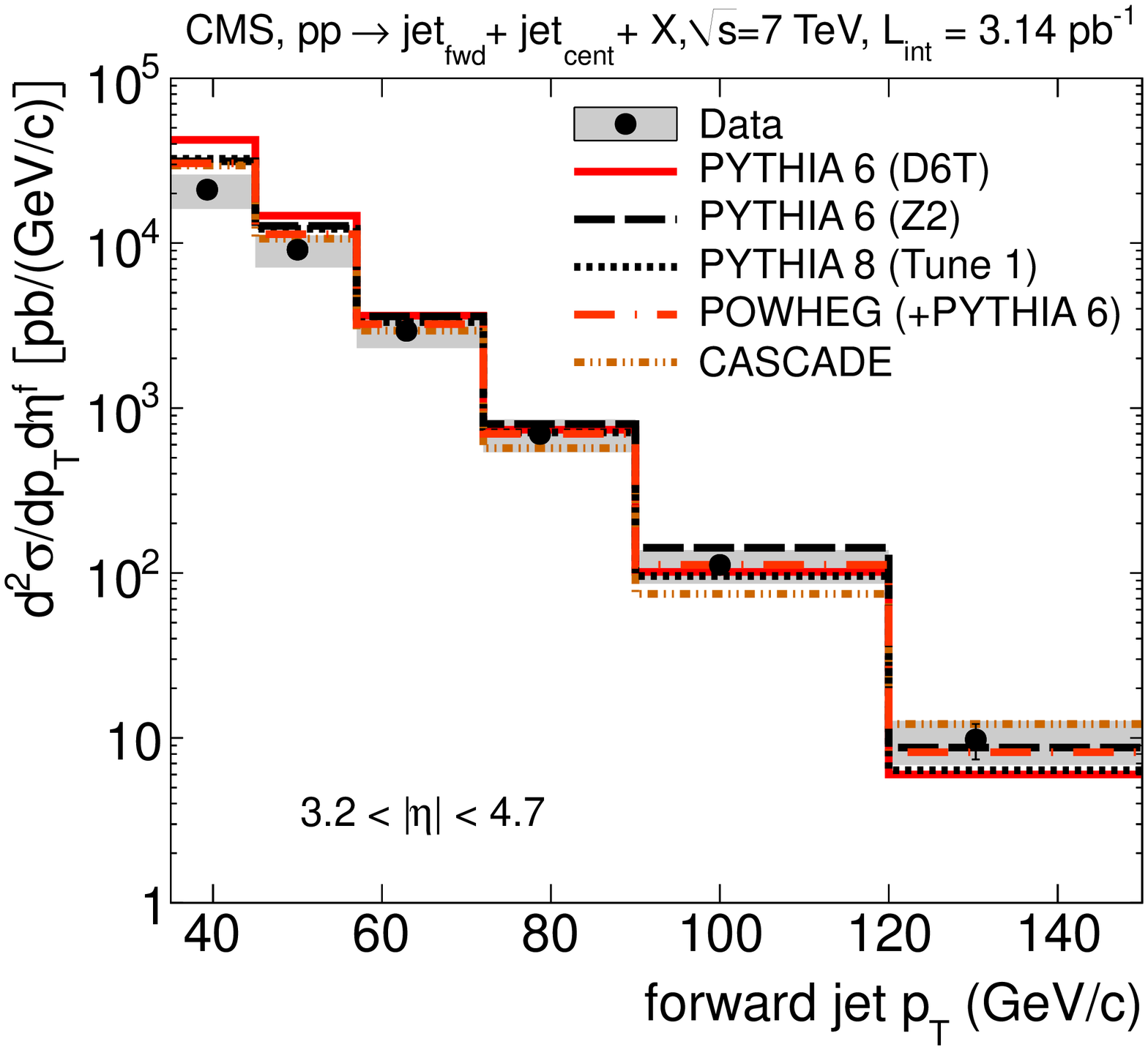,width=0.45\textwidth}\vspace{2mm}\\
  \epsfig{file=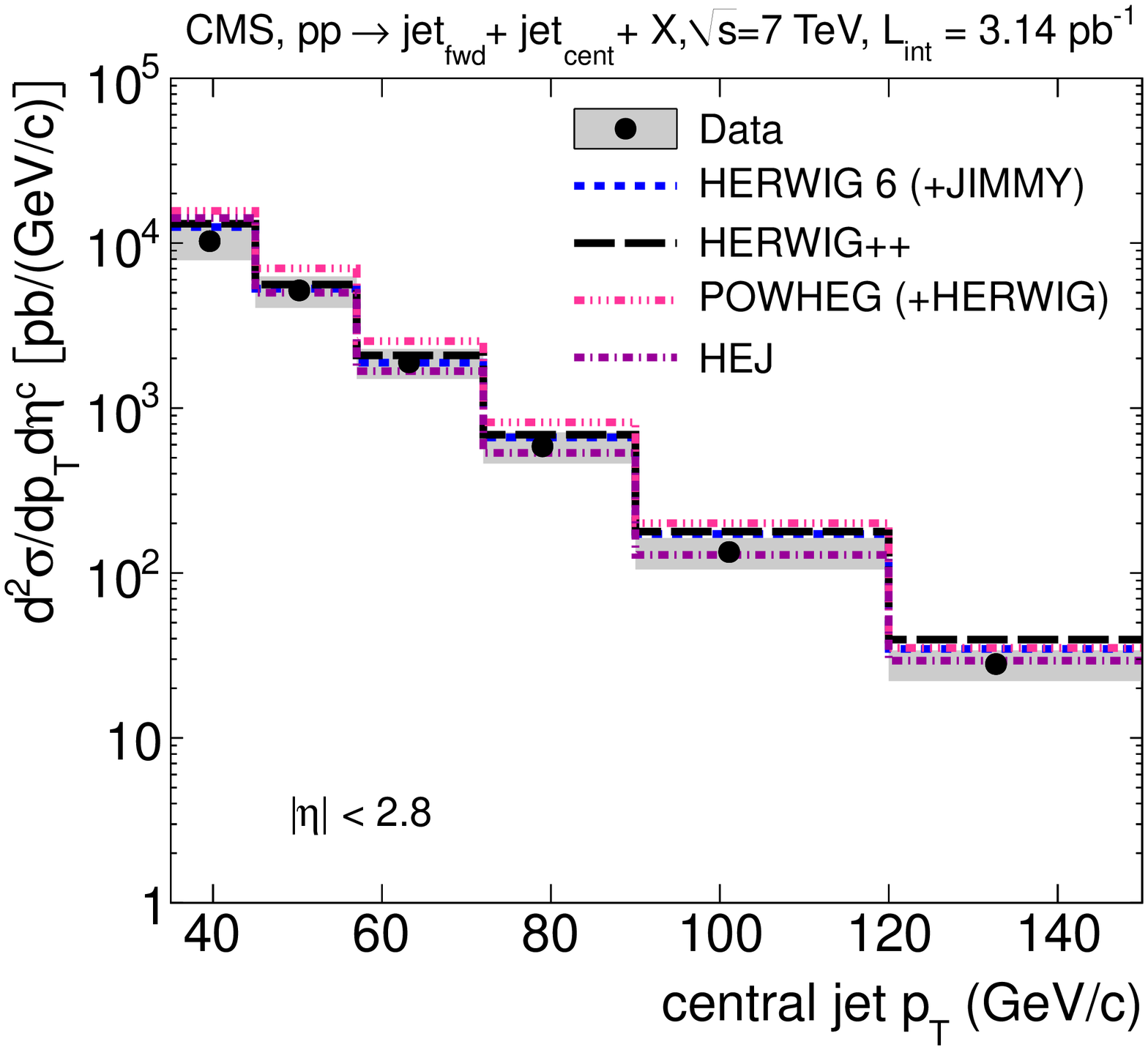,width=0.45\textwidth}\hfill
  \epsfig{file=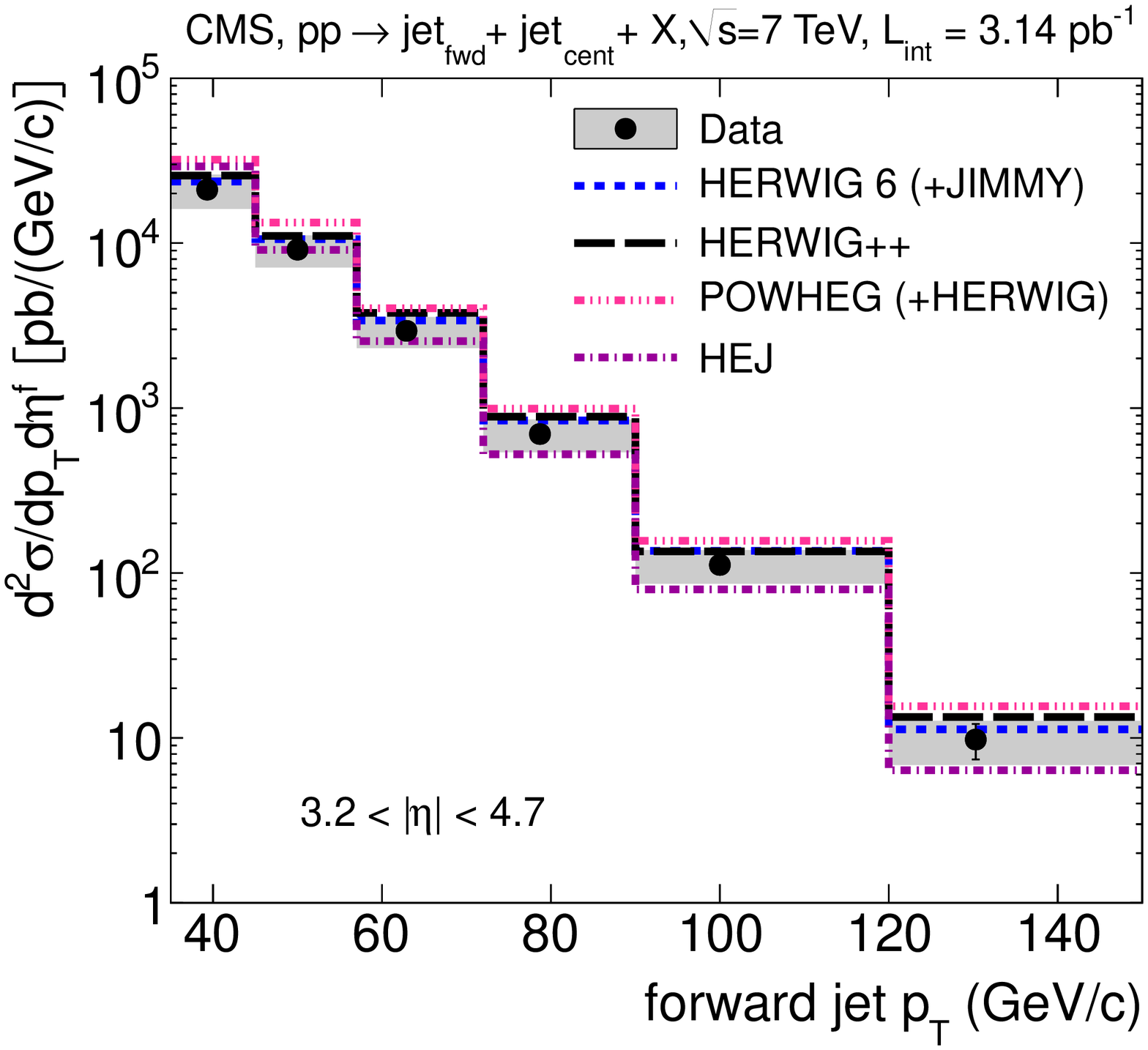,width=0.45\textwidth}
  \caption{The transverse-momentum distributions of the leading forward jet
    ($3.2<|\eta^f|<4.7$) and the leading central jet ($|\eta^c|<2.8$) in a
    sample which requires at least one jet with $\pT>35$~GeV in each
    region. The top and bottom row contain the same data points, but
    different theoretical predictions. The plots are taken
    from Ref.~\cite{Collaboration:2012gw}.}
  \label{fig:cms}
\end{figure}

CMS has reported a study~\cite{Collaboration:2012gw} on dijet production with
just a simple selection criteria on the transverse momenta of jets. Jets are
reconstructed by using the anti-$\kt$ algorithm with $R=0.5$, and are required to
have $\pT>35$~GeV.  Events are then required to contain at least one forward
jet ($3.2 <|\eta^f|<4.7$) and at least one central jet ($|\eta^c|<2.8$),
where $\eta$ is the pseudorapidity of the jets.  The transverse-momentum
spectrum of the hardest central and hardest forward jet is then
studied, see fig.~\ref{fig:cms}. Obviously, any difference in the two spectra is a result of
radiation beyond the tree-level description of back-to-back partons.
While the CMS study extends the pseudorapidity region of jets up to 4.7 units, the
transverse-momentum distributions are integrated over these pseudorapidity
ranges.
Crucially, however, no large $\pT$ hierarchy is induced by the cuts, and this
gives a cleaner study of the separate effects of the relatively modest
rapidity gap.

In the CMS analysis, \HEJ describes the $\pT$ spectrum well for both the
central and the forward jets. \PH describes the shape correctly,
but the normalization is consistently high.  The \PP description of the
forward jet $\pT$ distribution performs well, but the description for the
central jet shows deviations in both shape and normalization. As the events
in this analysis have been specifically selected to have a non-negligible
rapidity span, this slight deviation could be attributed to the absence
of a systematic treatment of the dominant logarithmic terms for increasing
$\Delta \eta$. 

\subsection{Summary}
The analyses discussed so far show that the descriptions of both \POWHEG and
\HEJ are performing well and in broad agreement. The close agreement between
the two can, to some extent, be attributed to the requirement of a large
$\pT$ hierarchy in the study by ATLAS or the modest average rapidity spans
in the study by CMS.

In the rest of this paper, we investigate various observables which can
expose the differences among the fixed NLO calculation, the \POWHEG and the \HEJ
approaches.  The first task is therefore to develop a set of cuts for which
the NLO prediction for dijet production is physically meaningful. This is the
topic of the next section.

\section{Reliability of the NLO predictions}
\label{sec:reliability-nlo}
It has been known for a while that fixed-order results for dijet production
are not reliable when symmetric cuts are applied to the transverse momentum
of the two hardest jets. This was first noticed in Refs.~\cite{Klasen:1995xe,
  Frixione:1997ks}, in the context of electron-proton collisions at HERA and
later also in hadron collisions~\cite{Andersen:2001kt}.  A detailed
theoretical discussion of the origin of this fact can be found
in Ref.~\cite{Banfi:2003jj}, where the next-to-leading-logarithmic resummation of
soft logarithms was also performed.

While the fixed-order theoretical predictions display an unphysical behavior
as the symmetric-cut limit is reached, the experimental data are obviously
not affected by that.  Since our goal in the following sections is to find
and discuss particular sets of cuts that will allow us to distinguish between
the two kinematic regimes implemented in the \POWHEGBOX{} and in \HEJ{}, and
since we would like to have a reliable NLO prediction to compare against,
in this section we study the reliability of the NLO differential cross
sections with several set of cuts, in order to find the most appropriate ones
to be used in the following comparisons.

We begin imposing the asymmetric cuts
\begin{equation}
  \label{eq:asymcuts}
  \pT^{\sss j} > \pT^{\min}\,,\qquad \pT^{\sss j_1} > \pT^{\min}\ + \Delta \pT\,, \qquad
  \Delta \pT > 0\,,
\end{equation}
i.e.~all jets are required to have a minimum transverse momentum
$\pT^{\min}$, while a stronger constraint is applied to the hardest jet in
the event, with transverse momentum $\pT^{\sss j_1}$. In our notation, $\Delta
\pT$ quantifies then the asymmetry on the cuts.  Since in this paper we are
interested in studying the size of QCD corrections in regions traditionally
used to probe BFKL-like effects, we allow for quite forward jets. Therefore
we impose the condition $|y_j|< 4.7$ on the jet rapidities, and we set
$\pT^{\min}=35$~GeV (similar to the CMS cut discussed in the previous
section).  Jets are reconstructed by using the anti-$\kt$ jet
algorithm~\cite{Cacciari:2008gp} with $R=0.5$ and $E-$scheme recombination.
  The NLO results, as well as the \POWHEG distributions shown in
section~\ref{sec:inclusive-variables}, have been obtained with 
renormalization and factorization scales set to the \POWHEG underlying-Born
transverse momentum, i.e.~the $\pt$ of the partons in the Born-like $2\to 2$
kinematics, the starting point for the generation of
radiation~\cite{Alioli:2010xa}. For all the plots in the paper we have used
the MSTW2008~\cite{Martin:2009iq} parton distribution function set.
The aim of this section is to explore several sets of cuts to be applied to
the jets and to clarify the breaking point, where the NLO calculation
becomes unreliable.

\begin{figure}[!btp]
  \centering
  \epsfig{file=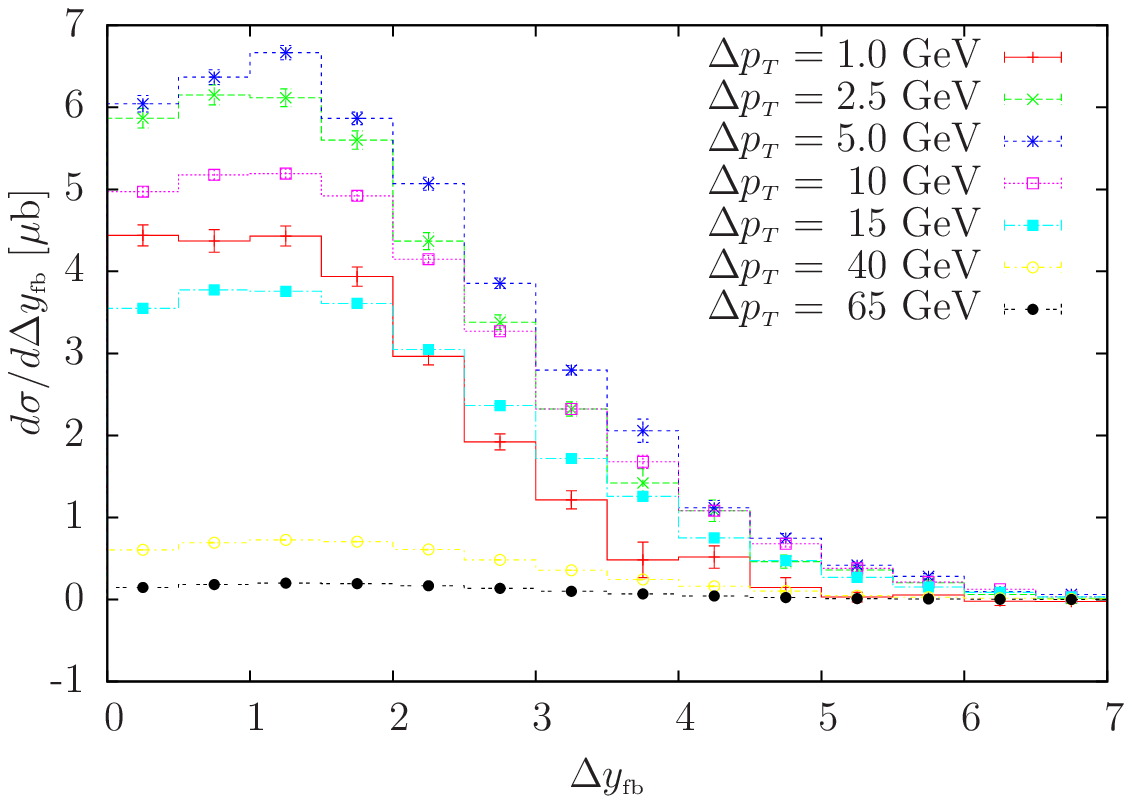,width=0.49\textwidth}
  \epsfig{file=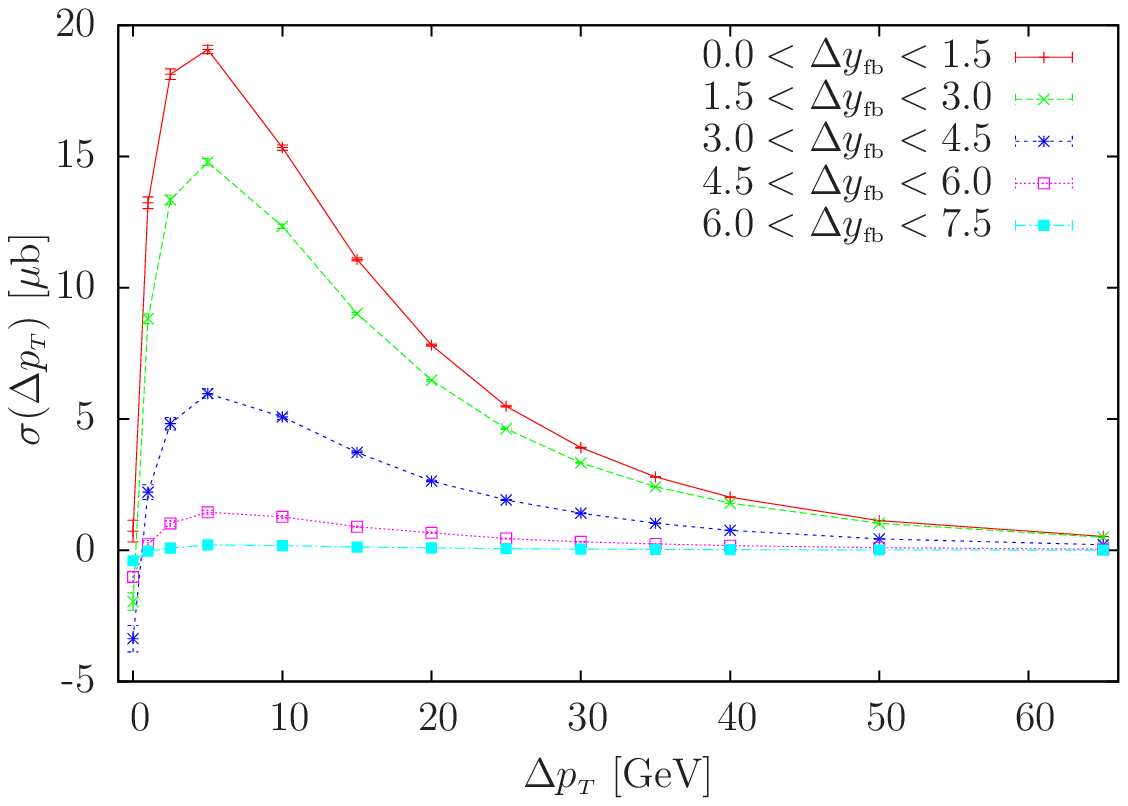,width=0.49\textwidth}
  \caption{The dependence of the $d\sigma / d\Delta y_{\sss \rm fb}$ distribution
    on the asymmetry of jet cuts (left plot) and of the total cross section
    in different $\Delta y_{\sss \rm fb}$ slices, as a function of the jet cuts
    asymmetry $\Delta \pt$ (right plot).}
  \label{fig:Dyjets_asymmetric}
\end{figure}
In fig.~\ref{fig:Dyjets_asymmetric}, we illustrate the effects of the cuts of
eq.~\eqref{eq:asymcuts}: in the left plot, we display the differential cross
section for dijet inclusive production as a function of $\Delta y_{\sss \rm
  fb}$, the difference in rapidity of the most forward and most backward jet.
The various curves correspond to different values of $\Delta \pT$ in
eq.~(\ref{eq:asymcuts}), ranging from $\Delta \pT=1$~GeV up to $\Delta
\pT=65$~GeV. 
On physical grounds, we expect a decrease of the differential cross
section as $\Delta \pT$ increases.  However
we observe that the NLO results do not show this behaviour. Indeed, over a
wide range in $\Delta y_{\sss \rm fb}$, the cross sections increase for
increasing $\Delta \pT$ until a maximum is reached at $\Delta \pT\approx
5$~GeV. Then, for any further increase of $\Delta \pT$, the cross section
decreases, as expected. The unphysical behaviour for small $\Delta \pT$ is
caused by a large logarithmic term in $\Delta \pt$, arising from a suppression
in the emission of radiation above the $\Delta \pt$ scale, which causes
uncancelled virtual corrections to build up above this scale.  The same
behaviour is also evident in the right plot of
fig.~\ref{fig:Dyjets_asymmetric}, where the cross section is plotted against
$\Delta \pT$ for various slices in $\Delta y_{\sss \rm fb}$. The presence of
uncanceled virtual corrections becomes manifest for small $\Delta \pT$, where
the cross section is unphysically negative.

\begin{figure}[!btp]
  \centering
  \epsfig{file=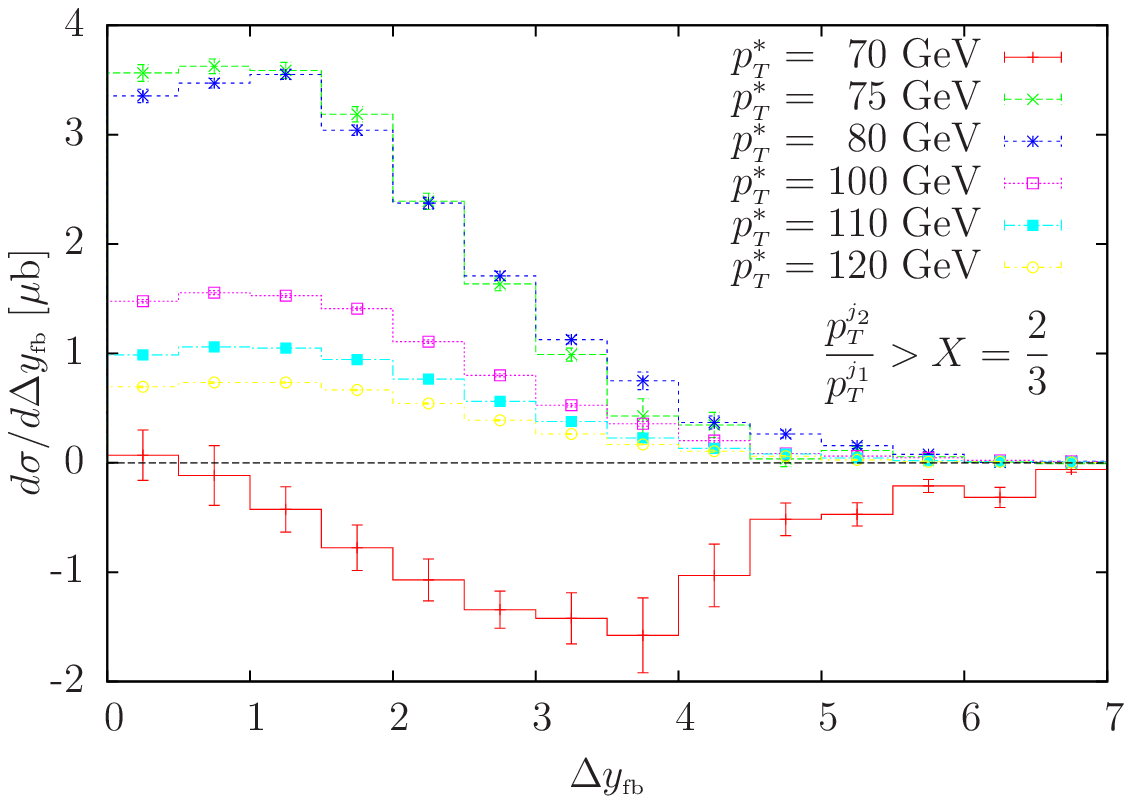,width=0.49\textwidth}
  \epsfig{file=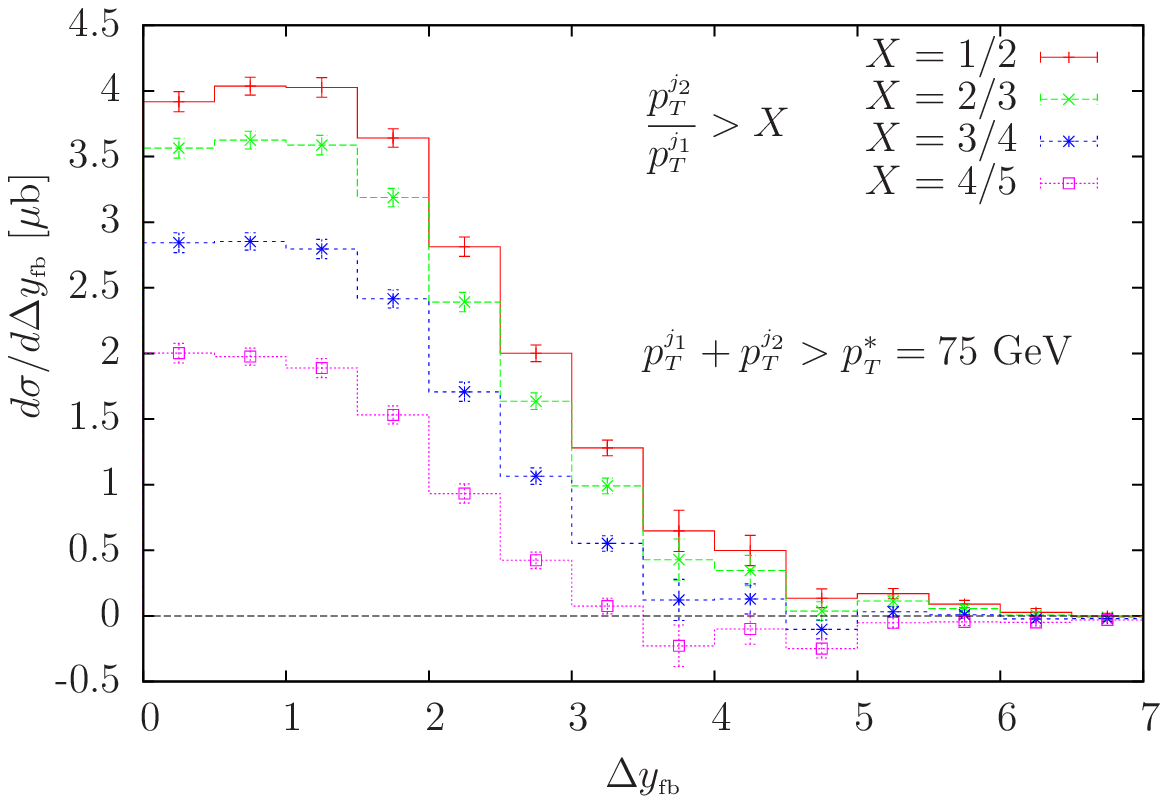,width=0.49\textwidth}
  \caption{The differential cross sections $d\sigma / d\Delta y_{ \sss \rm fb}$
    for several values of $\pt^*$ at $X=2/3$ (left plot) and for several
    values of $X$ for $\pt^*=75$~GeV (right plot). For both plots, jets have
  transverse momentum greater than 35~GeV.}
  \label{fig:Dyjets_sumscaling_Xscaling}
\end{figure}
In fig.~\ref{fig:Dyjets_sumscaling_Xscaling}, we plot the differential cross
section as a function of $\Delta y_{\sss \rm fb}$ using another set of cuts,
BFKL-inspired:
\begin{equation}
  \label{eq:sumcuts}
  \pT^{\sss j} > \pT^{\min}=35~{\rm GeV}\,, \qquad \pT^{\sss j_1} + \pt^{\sss j_2} > \pt^*\,,\qquad
  \frac{\pt^{\sss j_2}}{\pt^{\sss j_1}} > X\,,\qquad |y_{j}|< 4.7\,,
\end{equation}
where $j_1$ and $j_2$ denote the hardest and next-to-hardest jet, respectively.  In the
left plot of fig.~\ref{fig:Dyjets_sumscaling_Xscaling}, we show the cross
sections obtained at a fixed value of $X$ (chosen here to be $2/3$) for
several values of $\pt^*$~\footnote{While in principle $X$ can vary from $0$
  to $1$, it can be shown that, for a partonic $2\to n$ process, the lower
  limit is not 0, but $1/(n-1)$, so that, in our study ($n=3$), $1/2\le X \le
  1$.}.  The r\^ole of $\Delta\pt$ in fig.~\ref{fig:Dyjets_asymmetric} is now
played by the distance $\pt^*-2\,\pt^{\min}$. In fact, when $\pt^*$ reaches
its minimum value equal to $2\,\pt^{\min}$, jets can approach the
symmetric-cut configuration, exposing again large logarithms in a fixed NLO
calculation. The red curve shows precisely the unphysical behavior when
$\pt^*=2\,\pt^{\min}=70$~GeV, giving rise to a negative cross section.  We
have checked that this conclusion holds regardless of the value of $X$
chosen.  As $\pt^*$ increases beyond its minimal value, the cross sections
exhibit a physical behavior, i.e.~they decrease.
In the right plot of fig.~\ref{fig:Dyjets_sumscaling_Xscaling}, 
we display the differential cross sections for $\pt^*=75$~GeV and for several
values of $X$. The symmetric-cut regime is here reached when $X\to 1$, and in
fact, as the values of $X$ increases, the differential cross sections develop
negative tails for high $\Delta y_{\sss\rm fb}$.

\begin{figure}[!btp]
  \centering
  \epsfig{file=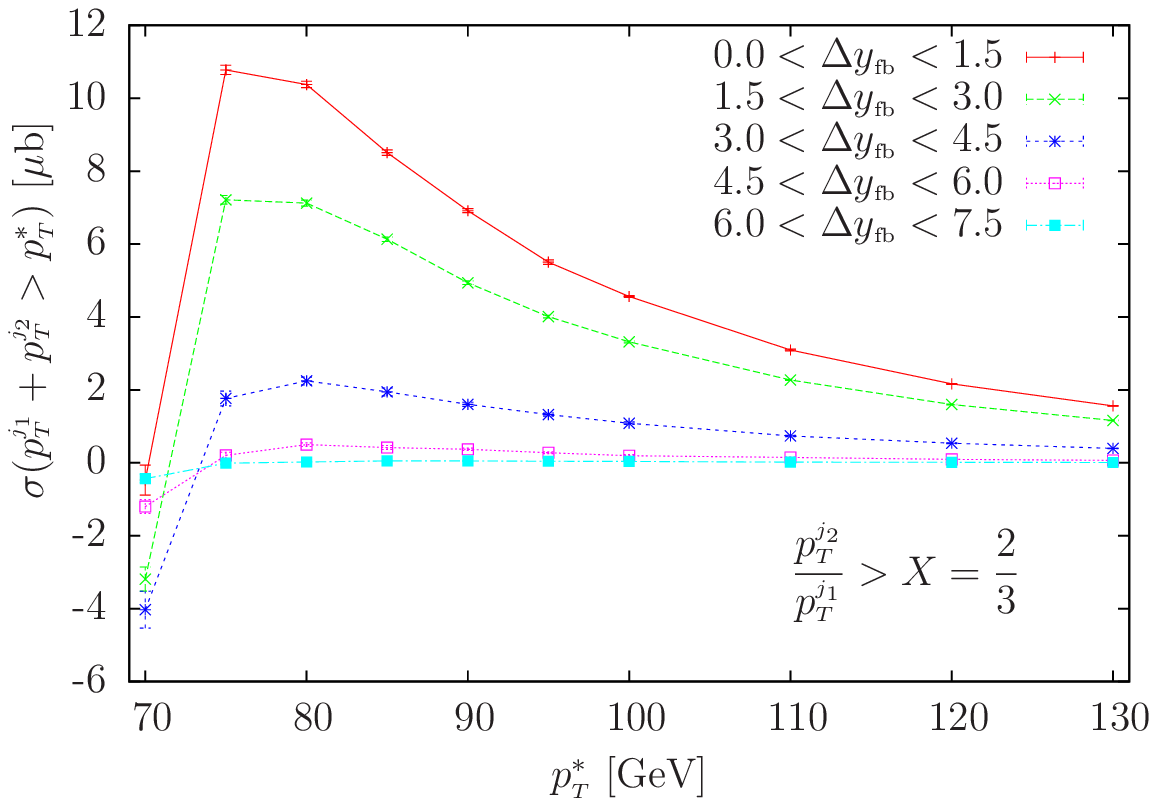,width=0.49\textwidth}
  \epsfig{file=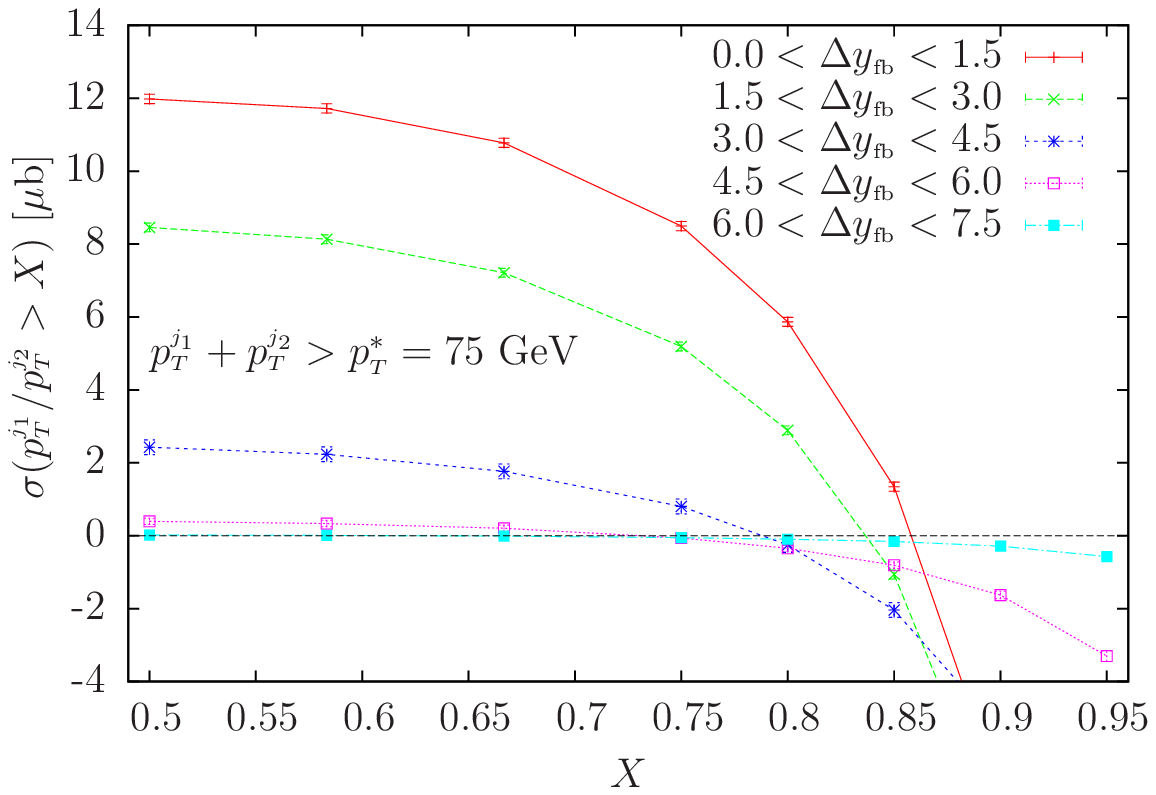,width=0.49\textwidth}
  \caption{Cross sections for different slices in $\Delta y_{\sss\rm fb}$ as
    a function of $\pt^*$ at $X=2/3$ (left plot) and as a function of $X$ at
    $\pt^*=75$~GeV (right plot).}
  \label{fig:Dyjets_sumscaling_Xscaling2}
\end{figure}
The two plots in fig.~\ref{fig:Dyjets_sumscaling_Xscaling2}, and similar ones
that can be drawn for other values of $X$ and $\pt^*$, can help in the
not-easy task of establishing what cuts to apply in an analysis if one wants
to compare the experimental data to the NLO calculation using the alternative cuts 
of eq.~\eqref{eq:sumcuts}.  In the
left plot, the cross sections plotted for several slices in $\Delta
y_{\sss\rm fb}$ become negative for values of $\pt^*$ less than 75~GeV,
fixing then a lower value below which the theoretical distributions cannot be
trusted.  Similarly, the right plot, where the cross sections are
plotted as a function of $X$ at a fixed value of $\pt^*=75$~GeV, gives an
indication of the upper value of $X$ above which the NLO predictions become
unreliable, for different slices of $\Delta y_{\sss\rm fb}$.

\section{Probing higher-order corrections: a comparison among \POWHEG, \HEJ and
NLO results}
\label{sec:inclusive-variables}

We are now in a position to compare three theoretical approaches to dijet
production that include higher-order effects: NLO, \POWHEG and \HEJ.  The aim
of this section is to investigate a number of observables based on hard jets,
which could better
expose the differences between the description obtained in the three approaches.

In order to avoid biasing our event sample towards a large hierarchy of
transverse scale, we impose a minimal set of asymmetric cuts (as discussed
around eq.~\ref{eq:goodcuts})
\begin{equation}
  \label{eq:goodcuts}
  \pt^{\sss j} > 35\ {\rm GeV},\quad \pt^{\sss j_1} > 45\ {\rm
    GeV},\quad |y_j| < 4.7\,,
\end{equation}
i.e.~all jets are required to have a minimum transverse momentum of 35~GeV,
and the hardest-jet transverse momentum, $\pt^{\sss j_1}$, is required to be
greater than 45~GeV. In order to comply with the experimental acceptance, all
jets are further required to have an absolute rapidity $|y_j|$ less than
4.7. Jets are defined according to the anti-$\kt$ jet algorithm, with radius
$R=0.5$. Only events with at least two jets fulfilling
eq.~(\ref{eq:goodcuts}) are kept. We stress that neither the \POWHEG nor the
\HEJ descriptions exhibit the unphysical behaviour of the NLO result when
using symmetric cuts, since they include a partial resummation of the large
logarithmic terms.  However, in order to have a meaningful NLO prediction to
compare with, we must impose asymmetric cuts.

In the following, we compare the cross sections computed with a fixed NLO
calculation and with \HEJ, with the results obtained analyzing 14M events
generated by the \POWHEGBOX, at the level of the first-emission and after the
shower performed by \PYTHIA~\footnote{We used \PYTHIA~6.4.25
  with the AMBT1 tune.}.  The renormalization and factorization
scales have been chosen equal to the transverse momentum of the hardest jet in
each event, for the \HEJ predictions.  For the NLO computation (and for
computing the \POWHEG $\bar{B}$ function), scales are set to the transverse
momentum of the underlying-Born configuration, as in the previous
section. Scale-uncertainty bands, obtained by varying these scales by a factor
of 2 in each direction, are shown for the NLO and \HEJ results.  The scales
entering in the evaluation of parton distribution functions and of the strong
coupling in the \POWHEG Sudakov form factor are instead evaluated with a scale
equal to the transverse momentum of the \POWHEG hardest
emission~\cite{Frixione:2007vw, Alioli:2010xa}.  

The statistical errors (due to the numerical integration of the differential
cross sections) on the theoretical predictions for the ratios studied in this
section have been computed with the standard propagation of errors for the
ratio of two uncorrelated quantities. We expect these errors to be an
overestimation of the true ones, since, in our case, the numerator and
denominator are strongly correlated.

\begin{figure}[btp]{ \centering
 \begin{minipage}[b]{.49\linewidth}
   \includegraphics[width=\textwidth]{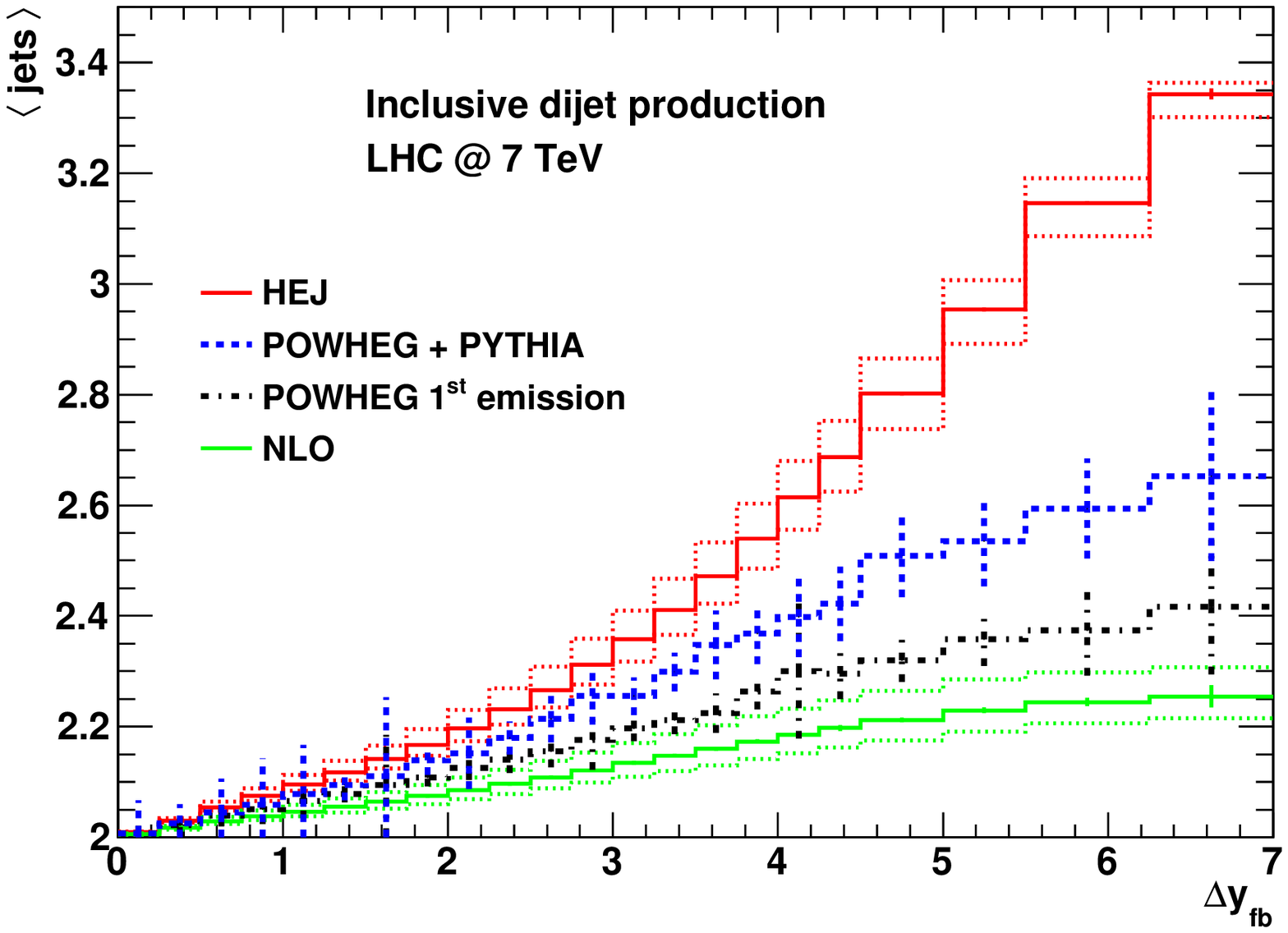}
   \includegraphics[width=\textwidth]{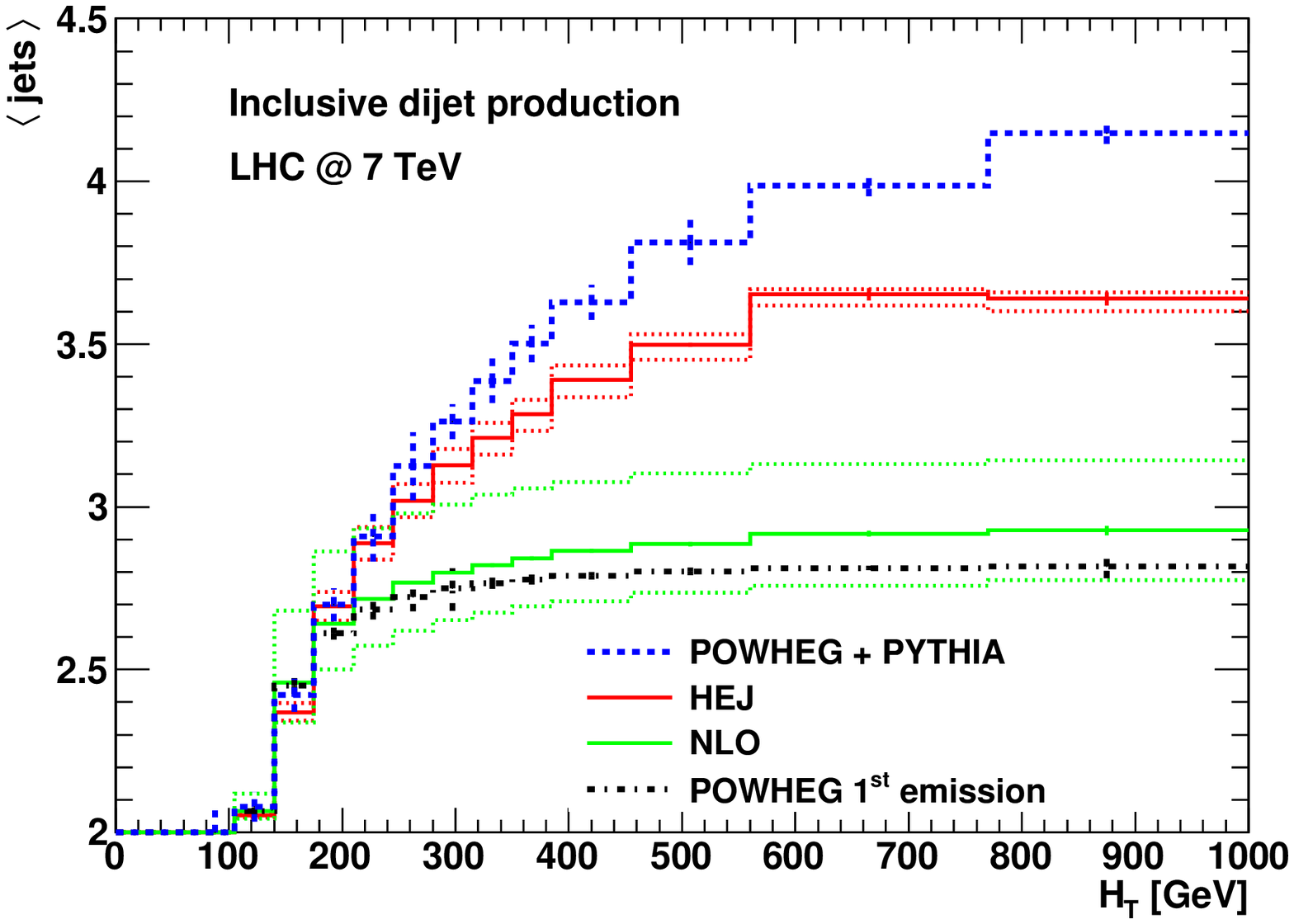}
 \end{minipage}
 \begin{minipage}[b]{.49\linewidth}
 \raisebox{-.053\height}{\includegraphics[width=0.95\textwidth]{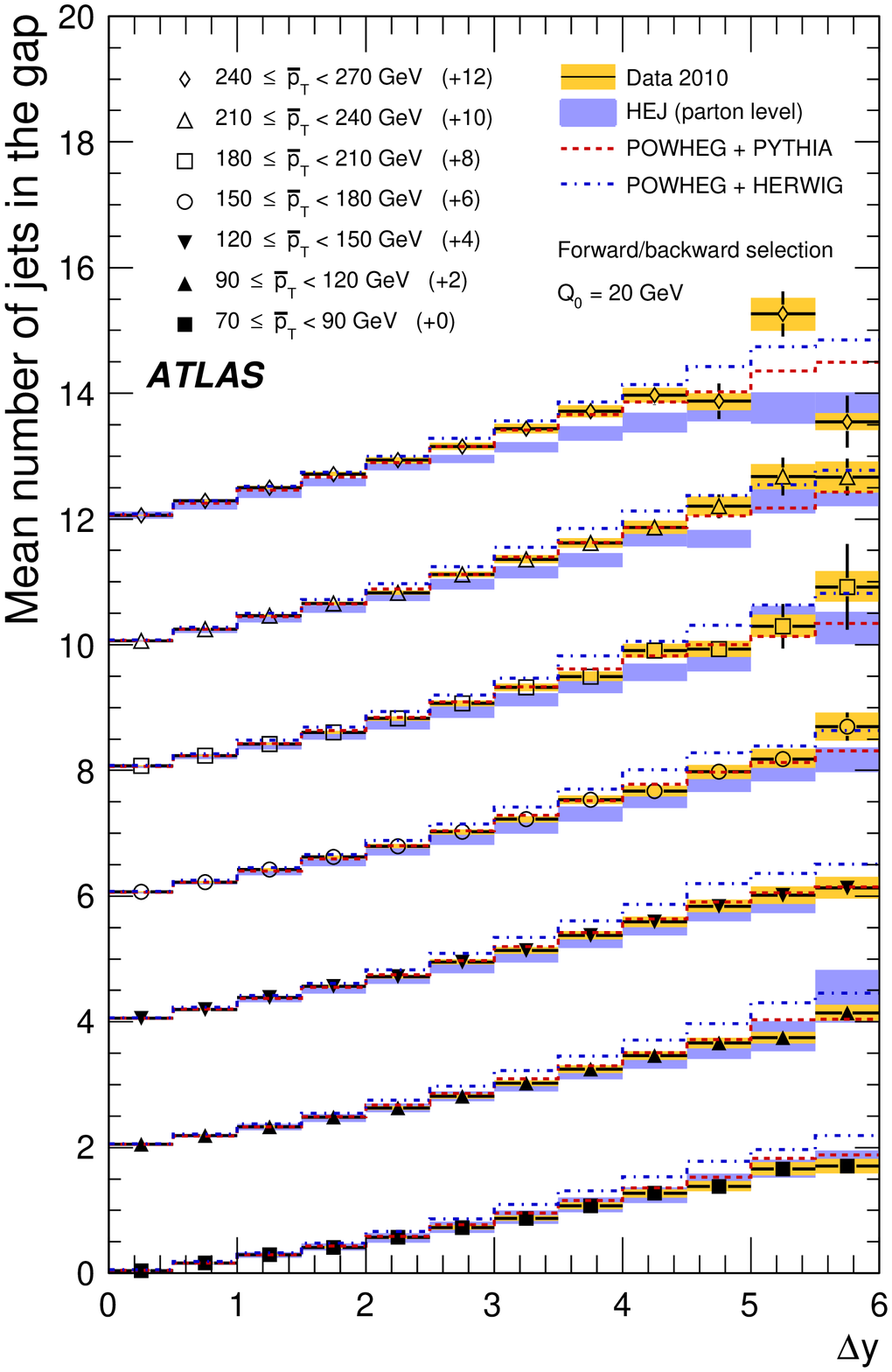}}
 \end{minipage}
 \caption{The average number of jets as a function of $\Delta y_{\rm fb}$
   (top left plot) and of $\HT$ (bottom left plot), as predicted by a fixed
   NLO calculation, by \POWHEG first emission, by \PP and by \HEJ.  The
   dotted red lines around the \HEJ prediction and the green ones around the
   NLO result are obtained by varying the renormalization and factorization
   scales by a factor of 2 around their central value.  The right plot (from
   Ref.~\cite{Aad:2011jz}) shows the average number of additional jets as
   found in the analysis from ATLAS.  }
 \label{fig:Dijets_avgjets}} \end{figure}
In fig.~\ref{fig:Dijets_avgjets}, we plot on the top left the average number
of jets as a function of the rapidity difference $\Delta y_{\sss \rm fb}$
between the most forward and most backward of the jets fulfilling
eq.~(\ref{eq:goodcuts}) and the same quantity as a function of $\HT=\sum_j
\pt^{\sss j}$ at the bottom left.  To ease the comparison, we also show the
result of the jet activity between the gap as a function of $\Delta y_{\sss
  \rm fb}$ in the analysis by ATLAS~\cite{Aad:2011jz} to the right of the
same figure. The analysis suggested in the present paper clearly shows more
discriminating power between the results of \POWHEG and \HEJ than the one of
Ref.~\cite{Aad:2011jz}.

As far as the dependence on $\Delta y_{\sss\rm fb}$ is concerned, the
wide-angle resummation implemented in \HEJ produces more hard jets than
\POWHEG and the fixed NLO calculation, as $\Delta y_{\sss \rm fb}$ increases.
Both the NLO and the first-emission \POWHEG results have at most 3 jets, so
that the average number of jets cannot exceed 3, and are in good
agreement. Additional jets are instead produced by the \PYTHIA shower, so
that the average number of jets is increased by roughly 20\% with respect to
the NLO one, for $\Delta y_{\rm fb}\approx 7$. For the same separation in
rapidity, the \HEJ prediction is 45\% larger than the NLO result, with a
chance to distinguish among the three approaches.  While this variable is
related to the gap fraction discussed previously, it is more exclusive as it
is sensitive to the number of jets in each event and not just whether a 3rd jet
exists.  We therefore anticipate greater distinguishing power between the
different theoretical approaches.  This same distribution has been
investigated also in the analysis by ATLAS~\cite{Aad:2011jz}, but, as
discussed earlier, the cuts applied in that analysis [in contrast to
eq.~(\ref{eq:goodcuts})] enhance the effects of collinear emissions, which
are treated to leading-logarithmic order in \POWHEG and partly in \HEJ.

The scale variation (dotted lines around the NLO and \HEJ curves) are modest, of
the order of a few percent.  As a final comment, we note that the prediction
from \HEJ was found to be very stable against the effects of further
showering~\cite{Andersen:2011zd} [by using cuts very similar to those in
eq.~(\ref{eq:goodcuts})].

\begin{figure}[btp]
  \centering
  \includegraphics[width=0.48\textwidth, bb = 12 5 525 450, clip]{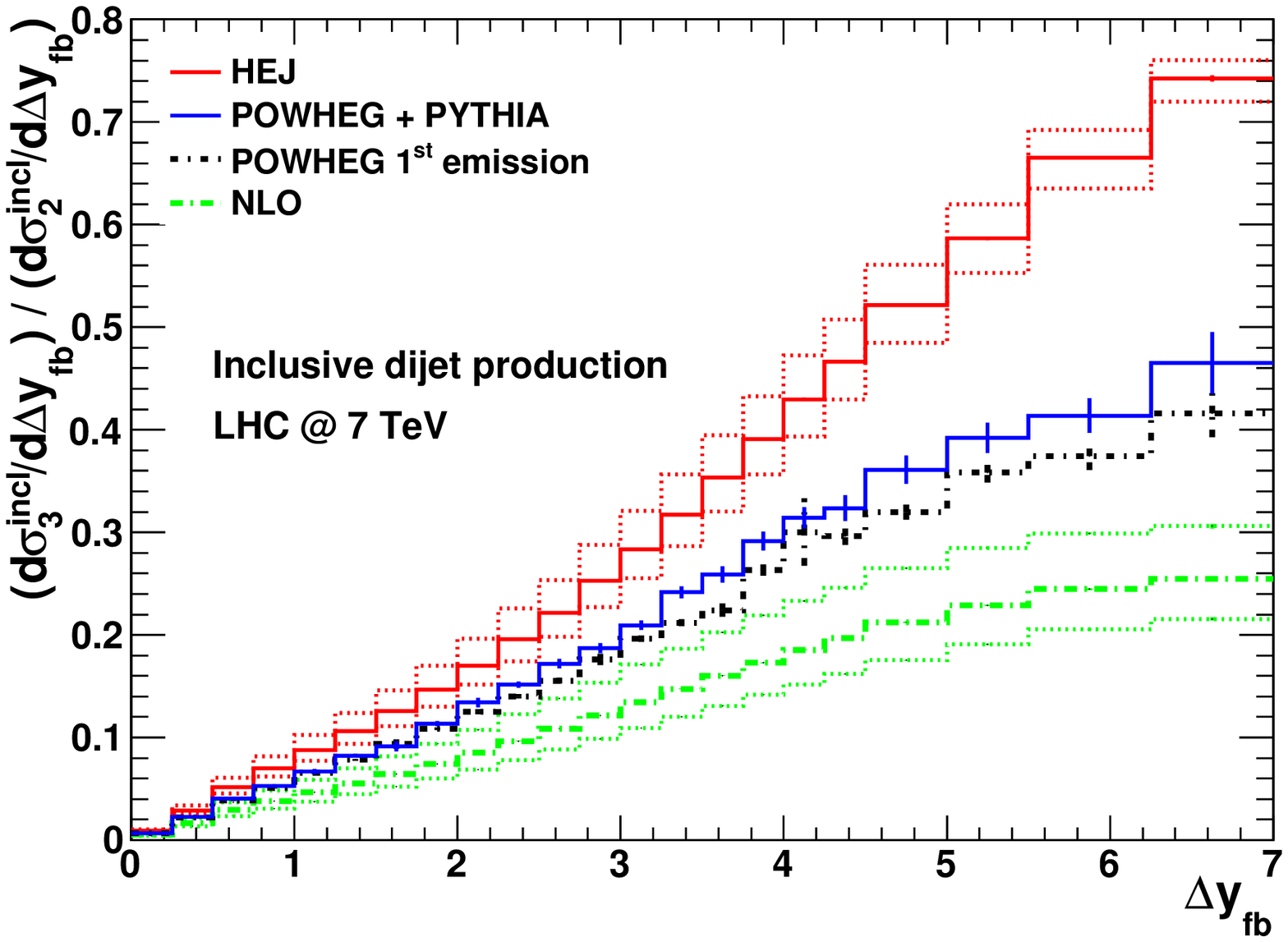}
  \includegraphics[width=0.48\textwidth, bb = 12 5 525 450, clip]{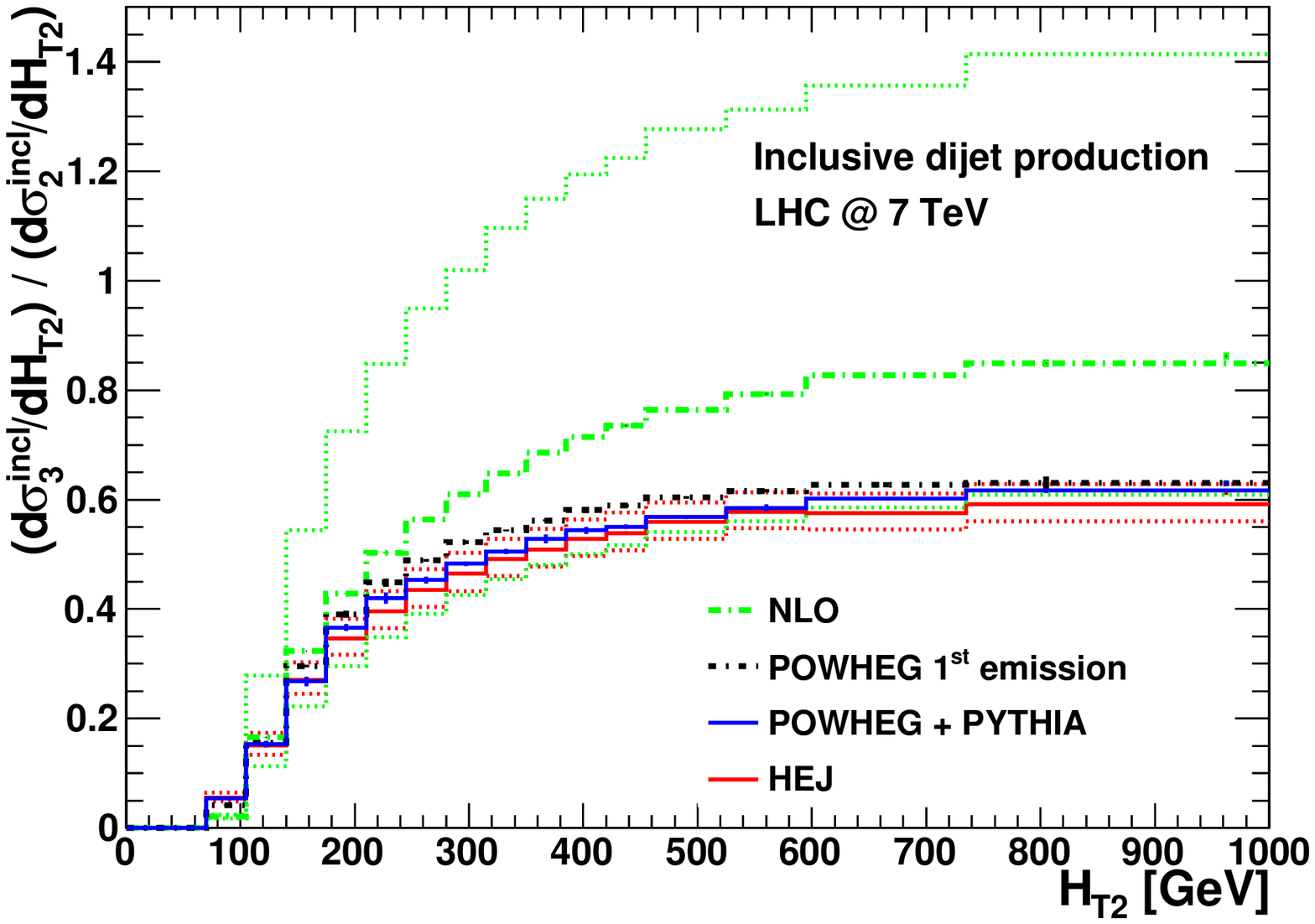}
  \caption{Ratio of the inclusive three-jet rate over the inclusive two-jet one, as
    a function of $\Delta y_{\sss\rm fb}$ (left plot) and $H_{\sss T2}$
    (right plot).  The dotted lines display the results obtained by varying
    the renormalization and factorization scales by a factor of 2 around
    the central value, for the \HEJ (red) and NLO (green) predictions.}
  \label{fig:incl3j2j}
\end{figure}

The dependence of the average number of jets on $\HT$ (bottom left plot in
fig.~\ref{fig:Dijets_avgjets}) displays a different behaviour: here the
showered events have, on average, more jets than \HEJ and the NLO results, as
the sum of the transverse momentum of all the final-state jets increases.
It is interesting here to comment on the NLO result obtained with the
factorization and renormalization scales set to $\pt^{\rm \scriptscriptstyle
  UB}/2$, half of the transverse momentum of the underlying-Born
configuration, i.e.~the upper green dotted line in the plot.  In fact, this
quantity is greater than 3, for $\HT\gtrsim 270$~GeV, which in a NLO
calculation signals the fact that the two-jet exclusive cross section becomes
negative.  We will comment on this after the discussion of
fig.~\ref{fig:incl3j2j} that suffers from the same problem.

In fig.~\ref{fig:incl3j2j}, we plot the ratio of the three-jet inclusive cross
section over the two-jet one, as a function of $\Delta y_{\rm \sss fb}$ (left
plot) and as a function of $H_{\sss T2}$ (right plot), where $H_{\sss
  T2}=\pt^{\sss j_1} + \pt^{\sss j_2} $ is the sum of the transverse momenta
of the two hardest jets in the event.  The same comments made about the top
left plot of fig.~\ref{fig:Dijets_avgjets} apply here: the BFKL-inspired
resummation implemented by \HEJ produces more hard jets than the resummation
of the parton shower of \PP, for large rapidity separation between the most
forward and most backward jet.  For this distribution, an experimental
analysis should then be able to distinguish between the \HEJ and the \PP
predictions, even though the observable is less exclusive than the average
number of jets (and directly related to the gap fraction studied by ATLAS).

In the right plot, the NLO ratio for the three-jet inclusive cross section over
the two-jet one, plotted as a function of $H_{\sss T2}$, becomes unphysical
(i.e.~it becomes greater than 1) when the factorization and renormalization
scales are set to $\pt^{\rm\sss UB}/2$, for higher values of $H_{\sss T2}$.
This result is linked with the same unreliable behavior of the NLO
distribution in the bottom left plot of fig.~\ref{fig:Dijets_avgjets} and
deserves an explanation that we give in Appendix~\ref{app:sig2excl}. The
predictions from \HEJ and \PP are in remarkably good agreement, both close to
the \POWHEG first-emission result, implying that the first \POWHEG emission
has the strongest impact on this distribution, while the subsequent shower
has a milder effect.  Before leaving this discussion, we would like to point
out that, when using the cuts reported by the ATLAS
Collaboration~\cite{Aad:2011tqa}
\begin{equation}
  \pt^{\sss j} > 60\ {\rm GeV},\quad \pt^{\sss j_1} > 80\ {\rm
    GeV},\quad |y_j| < 2.8\,,
\end{equation}
no significant difference is generated between the NLO result and the other
three curves (in complete agreement with the results reported in Ref.~\cite{Aad:2011tqa}).

\begin{figure}[!btp]
  \centering
  \epsfig{file=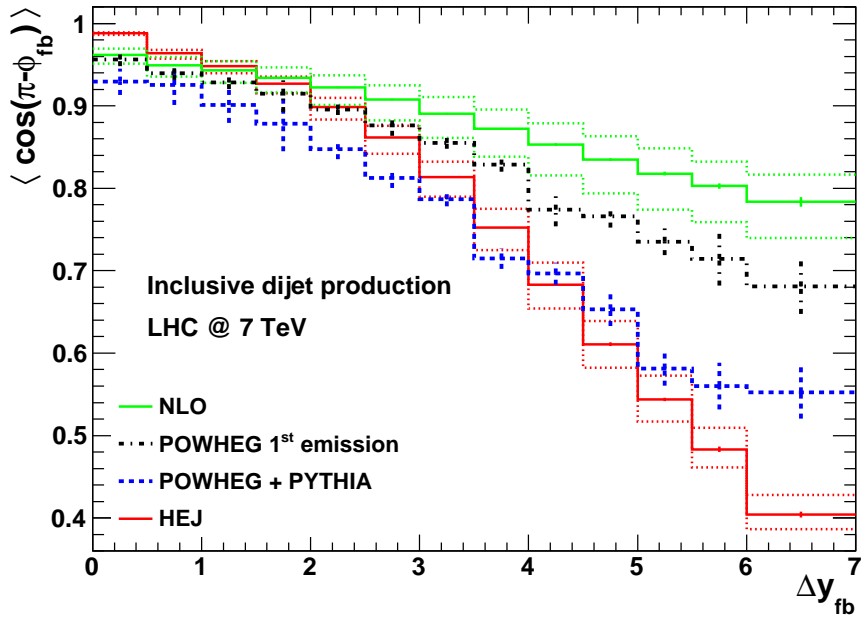,width=0.7\textwidth}
  \caption{The average value of $\cos(\pi-\phi_{\rm\sss fb})$ as a function
    of $\Delta y_{\rm \sss fb}$, where $\phi_{\rm\sss fb}$ is the azimuthal
    angle separation between the most forward and most backward jet.  The
    dotted red and green lines are obtained by varying the renormalization
    and factorization scales by a factor of 2 in both directions around the
    central value.}
  \label{fig:cosphi}
\end{figure}
As a last example of a kinematic distribution that displays different behaviour
if evaluated at NLO or by using \POWHEG or \HEJ, we plot, in
fig.~\ref{fig:cosphi}, the average value of $\cos(\pi-\phi_{\rm \sss fb})$,
where $\phi_{\rm \sss fb}$ is the azimuthal angle between the most forward
and backward jets, as a function of their rapidity separation $\Delta y_{\rm
  \sss fb}$. For dijet events at tree level, $\phi_{\rm \sss fb}=\pi$, since
the two jets are back-to-back, and the average value of the cosine is 1.
Deviation from 1 then indicates the presence of additional emissions, so that
this kinematic distribution carries information on the decorrelation between
the two jets.  This observable has been promoted for a long time as a good
discriminator between descriptions with and without a systematic evolution in
rapidity. It has therefore also been studied with a full detector simulation in
ref.~\cite{d'Enterria:2009fa}. However, the striking prediction from pure
leading-logarithmic BFKL evolution of an azimuthal decorrelation much larger
than that which is obtained in a parton shower or fixed-order formulation has
been brought into question for some time by the inclusion of subleading
corrections~\cite{Orr:1997im,Colferai:2010wu}.
This quantity is more inclusive than the average number of jets, as it is
sensitive also to emissions below the jet $\pt$ cut.  The higher radiation
activity in \PP and in \HEJ, with respect to the fixed NLO and
the \POWHEG first-emission results, is clearly visible in the figure: the
stronger jet activity produced by \HEJ at higher rapidity separation (see the
left plot of fig.~\ref{fig:Dijets_avgjets}) lowers the average value of the
cosine below the \PP result.  As expected, the average value
predicted by the \POWHEG first-emission and the NLO calculation is closer to
1, since they contain at most one radiated parton. At large rapidity
separations, the prediction from dijets at NLO is of significantly less
decorelation than that from either \PP or \HEJ.

Concluding the study of distributions we note that when the average number of
jets is analysed vs~the hardness of the event measured as $H_T$ and the
ordering of predictions from fewer to more hard jets is NLO, \HEJ, \PP. In
the more inclusive analysis of the three-jet rate over the two-jet rate
vs~$H_{T2}$, the results of \PP and \HEJ are very similar, and the NLO
prediction is of just a slightly larger share of inclusive three-jet events.

However, when the additional jet activity is studied as a function of
the rapidity separation between the most forward/backward jets, then
the results of the perturbative predictions are systematically ordered
from least to most radiation as: NLO, \PP and \HEJ. This is seen in
fig.~\ref{fig:Dijets_avgjets} for the average number of jets and in
fig.~\ref{fig:incl3j2j} for the inclusive three-jet rate over the
inclusive two-jet rate. We can thus confirm that the a priori expected
behaviour between the various perturbative frameworks is indeed
realized in practice.

\section{Conclusions}
\label{sec:conclusions}
Recent analyses by the ATLAS and the CMS Collaborations of inclusive and
exclusive dijet production showed a high level of agreement between the two
very different approaches to the description of perturbative higher-order
corrections implemented in the \POWHEGBOX and in \HEJ, within the specific
cuts and analyses applied.

Inspired by these results, we have presented an analysis developed to clearly
display the differences in the radiation patterns arising in a fixed NLO
calculation, \HEJ and \PP. All the observables discussed probe directly the
further radiation from a dijet system, so the NLO calculation of dijet
production is the lowest order nontrivial prediction for the observable.

While the limitations of the NLO calculation are clearly visible when
probing regions of the phase space where multijet emissions become
important, we have shown also that the predictions of \PP and \HEJ are
clearly distinguishable for the average number of jets and the ratio
of the inclusive three-jet production over the inclusive two-jet
production, when studied as a function of the rapidity separation of
the most forward and the most backward jet.  Less marked differences
are found when these quantities are plotted as a function of the sum
of the transverse momenta of all the jets, or as a function of the
transverse momenta of the two leading jets.  Contrary to these
findings, the study of the azimuthal decorrelation of the most forward
and backward jet turned out to be less promising in distinguishing the
two descriptions given by \POWHEG and \HEJ{} - while effects beyond
pure NLO should be clearly visible.

We hope that an experimental measurement of dijet data collected at
the LHC based on the suggestions presented in this paper will follow,
in order to investigate the quality of the theoretical understanding
of these kinematic distributions.

\section*{Acknowledgments}
\label{sec:acknowledgements}

The authors thank the staff of the Ecole de Physique des Houches for their
hospitality and the organizers of the ``Physics at TeV Colliders 2011''
workshop held there, where this project was started.  We are also grateful
for useful discussions with Gavin Salam and the other members of the LPCC
Small-$x$ discussion forum on a number of occasions.  ER and JMS are
supported by the UK Science and Technology Facilities Council (STFC).  ER and
SA acknowledge financial support from the LHCPhenoNet network under Grant
Agreement No.~PITN-GA-2010-264564 for travel expenses.

\appendix
\section{The exclusive two-jet cross section}
\label{app:sig2excl}
In this appendix we give an explanation of the unphysical behavior of the NLO
distributions shown in the bottom left plot of fig.~\ref{fig:Dijets_avgjets}
and in the right plot in fig.~\ref{fig:incl3j2j}, when the factorization and
renormalization scales are set to $\pt^{\rm\sss UB}/2$.  For ease of notation
we introduce the following shortcuts
\begin{equation}
  \sigdi = \frac{d \sigma_2^{\rm \sss incl}}{d H_{\sss T2}}\,, \qquad \quad
  \sigti = \frac{d \sigma_3^{\rm \sss incl}}{d H_{\sss T2}}\,,
\end{equation}
where the lower index 2 or 3 indicates the number of jets and the
differential cross sections are inclusive with respect to the corresponding
number of jets. Together with inclusive cross sections, we define the
exclusive ones, that will be designated with the upper label ``excl''.
At fixed NLO, in dijet production, we have
\begin{equation}
\sigti = \sigte\,, \qquad  \quad \sigdi = \sigde + \sigte\,,
\end{equation}
and we can relate the average number of jets with the inclusive three-jet over
two-jet ratio
\begin{equation} 
\langle {\rm jets}\rangle \equiv \frac{2\,\sigde + 3\,\sigte} {\sigde+ \sigte} = 2
+ \frac{\sigte}{\sigdi} = 2 + \frac{\sigti}{\sigdi}
\end{equation}
so that the two unphysical behaviors of the right plots in
figs.~\ref{fig:Dijets_avgjets} and~\ref{fig:incl3j2j} are strictly connected.
The ratio 
\begin{equation} 
\frac{\sigti}{\sigdi} =\frac{\sigti}{\sigde+\sigti}
\end{equation}
can become greater than 1 for particular kinematic configurations only if the
exclusive two-jet cross section becomes negative at those phase space
points. This happens, in our plots, when we choose $\pt^{\rm \sss UB}/2$ as
factorization and renormalization scale, for values of $H_{\sss T2} \gtrsim
270$~GeV.  We have explicitly checked that the same behavior is observed
if one sets the scale to be the hardest transverse momentum of the NLO partonic
kinematics, a scale that is generally used for this kind of process.

\begin{figure}[htp]
  \centering
  \epsfig{file=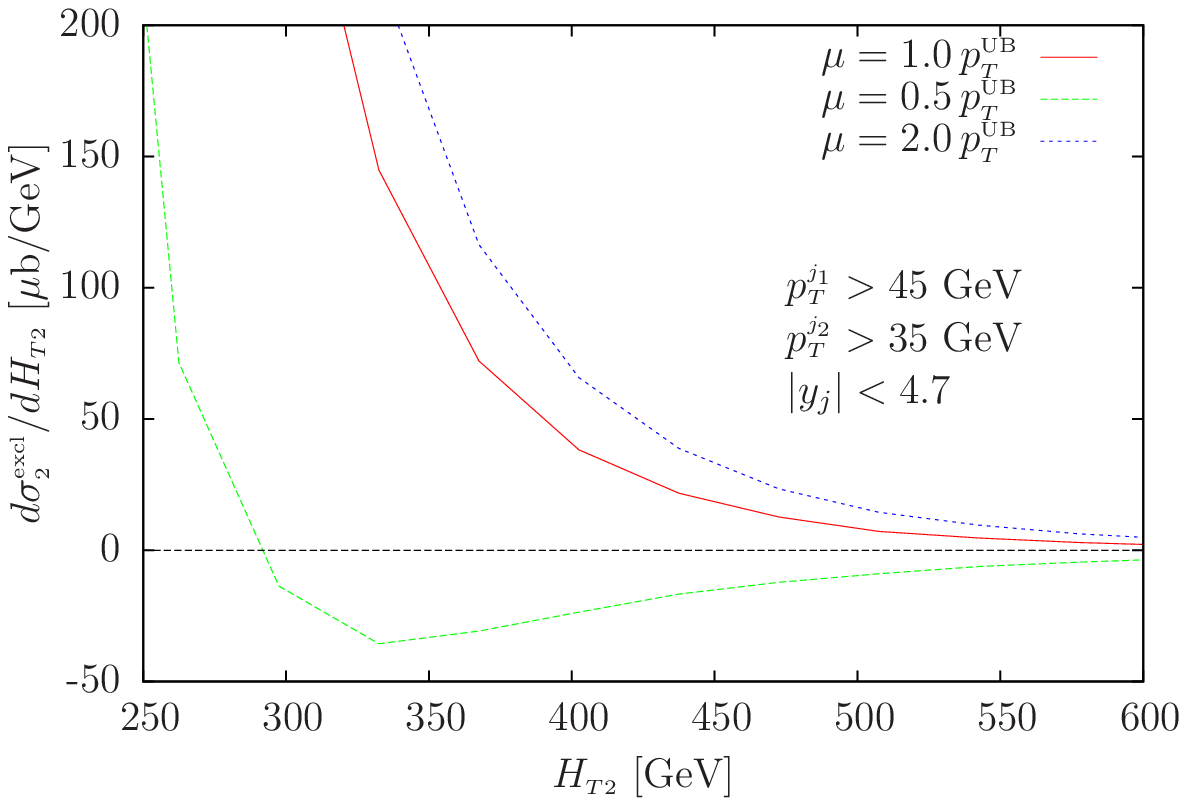,width=0.49\textwidth}\hfill
  \epsfig{file=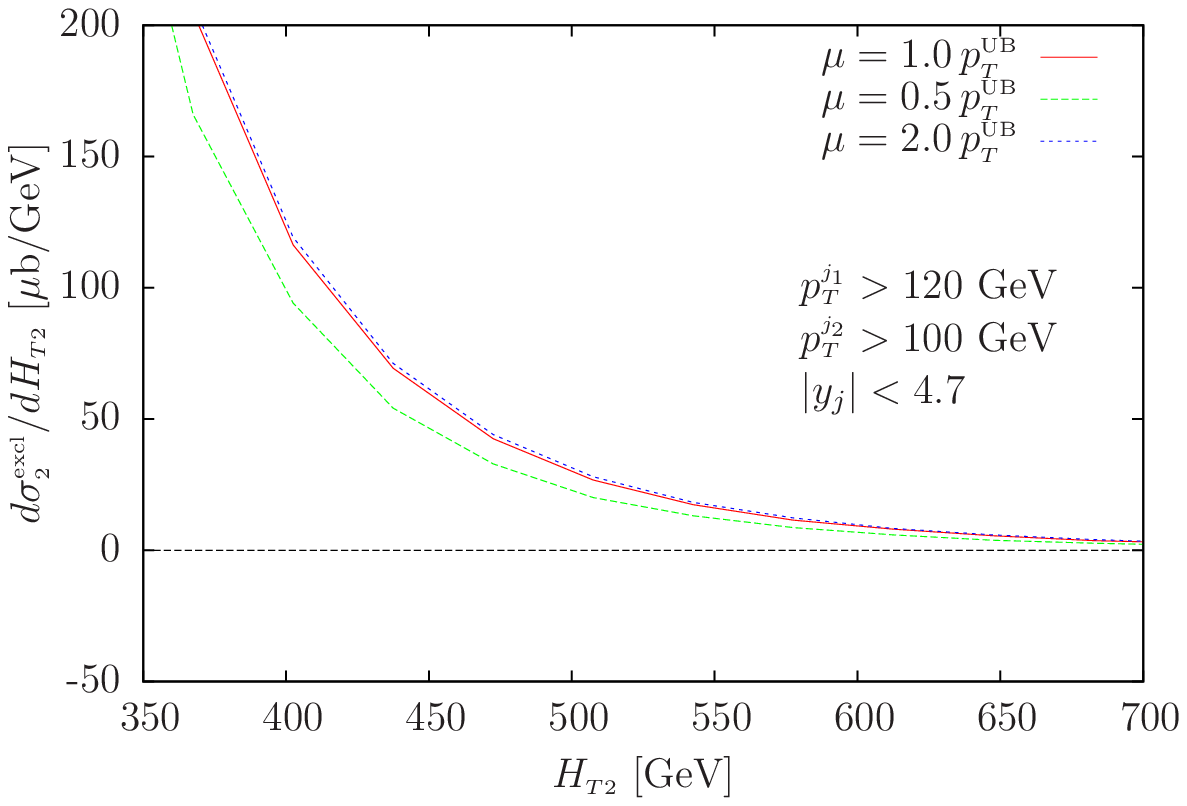,width=0.49\textwidth}
  \caption{Exclusive two-jet cross section as a function of $H_{\sss T2}$, for
    three values of the renormalization and factorization scale $\mu$ $\lg
    \pt^{\sss\rm UB}, \pt^{\sss\rm UB}/2, 2\,\pt^{\sss\rm UB}\rg$ and for
    different sets of cuts on the minimum-jet $\pt$.  }
  \label{fig:sigdexcl}
\end{figure}
In the left plot of fig.~\ref{fig:sigdexcl}, we plotted $\sigde$, and, as
expected, it becomes unphysical when the scale chosen is $\pt^{\rm \sss
  UB}/2$ and for values of $H_{\sss T2} \gtrsim 270$~GeV.  The explanation of
this can be again traced back to the large logarithmic terms related to
symmetric cuts and to the increase of the value of $\as$, now evaluated at a
smaller scale (the r\^ole played by the factorization scale would be more
difficult to disentangle, since it involves the behavior of the parton
distribution functions too).  The two-jet exclusive cross section always gets
a contribution from the Born and the virtual terms, irrespective of
the value of $H_{\sss T2}$, and from the part of the real-emission cross
section that, at those kinematic points, is clustered into a two-jet
configuration
\begin{equation} 
\sigde = \tilde{\sigma}_2^{\sss B} + \tilde{\sigma}_2^{\sss V} +
\tilde{\sigma}_2^{\sss R} = \as^2(\mu)\big\{ B + \as(\mu) \big[ V + R_2 \big]\big\} \,, 
\end{equation}
where the notation is self-explanatory and we put in evidence the appearance of
the strong coupling constant $\as(\mu)$ evaluated at the scale $\mu$.  Since $B$
and $R_2$ are necessarily positive, coming from the square of the respective
matrix element, it is the virtual term $V$ that drives $\sigde$ to negative
values. In other words, the $R_2$ term becomes smaller and smaller if compared
to the absolute value of $V$.  At high values of $H_{\sss T2}$ , most of the
events have three jets, with the two hard jets with $\pt\sim H_{\sss T2}/2$,
because, most likely, the transverse momentum of the third jet is just high
enough to pass the cuts in eq.~(\ref{eq:goodcuts}). Since the minimum $\pt$ for
the jets is 35~GeV, the imbalance between the two hardest jets is small, and the
situation is equivalent to imposing symmetric cuts.  On the other hand, if we
increase to 100~GeV the minimum transverse momentum to define a jet, as in the
right plot of fig.~\ref{fig:sigdexcl}, the imbalance between the two hardest jets
is no longer small compared to their average transverse momentum, and the event
kinematics stays away from the symmetric-cut configuration.
In addition, we have explicitly checked that, by keeping the $\pt$ cuts of
eq.~(\ref{eq:goodcuts}) but restricting the jet-rapidity range to $|y_j|<
2.5$, the critical $H_{\sss T2}$ value is moved to higher values,
i.e.~$\sim \! 400$~GeV, implying that the unphysical behaviour depends on 
both the transverse momentum and the rapidity cuts.

\bibliography{papers}

\end{document}